\documentclass[12pt,preprint]{aastex}
\usepackage{graphicx}
\usepackage{natbib} 
\shorttitle{Internal rotation of the red-giant star KIC~4448777}
\shortauthors{Di Mauro et al.}
\begin{document}
\title{Internal rotation of the red-giant star KIC~4448777 by means of asteroseismic inversion}
\author{M. P. Di Mauro\altaffilmark{1}, R. Ventura\altaffilmark{2}, D. Cardini\altaffilmark{1}, D. Stello\altaffilmark{3,4},
J.~Christensen-Dalsgaard\altaffilmark{4}, W.~ A.~Dziembowski\altaffilmark{5,6}, L. Patern\`o\altaffilmark{2}, P. G. Beck\altaffilmark{7,8}, S. Bloemen\altaffilmark{9}, G.~R.~Davies\altaffilmark{8,10}, K.~De Smedt\altaffilmark{7}, Y. Elsworth\altaffilmark{10}, R. A. Garc\'ia\altaffilmark{8}, S. Hekker\altaffilmark{4,11,12}, B. Mosser\altaffilmark{13}, and A.~Tkachenko\altaffilmark{7}}
\altaffiltext{1}{INAF, IAPS Istituto di Astrofisica e Planetologia Spaziali, Roma, Italy}
\altaffiltext{2}{INAF, Astrophysical Observatory of Catania, Catania, Italy}
\altaffiltext{3}{Sydney Institute for Astronomy, School of Physics, University of Sydney, Australia}
\altaffiltext{4} {Stellar Astrophysics Centre, Department of Physics and Astronomy, Aarhus University, Ny Munkegade 120, DK-8000 Aarhus C, Denmark}
\altaffiltext{5} {Warsaw University Observatory, Al. Ujazdowskie 4, 00-478 Warsaw, Poland}
\altaffiltext{6} {Copernicus Astronomical Center, ul. Bartycka 18, 00-716 Warsaw, Poland}
\altaffiltext{7}{Instituut voor Sterrenkunde, Katholieke Universiteit Leuven, Belgium}
\altaffiltext{8}{Laboratoire AIM, CEA/DSM-CNRS-Univ. Paris Diderot, IRFU/Sap, Centre de Saclay, 91191 Gif-sur-Yvette Cedex, France}
\altaffiltext{9}{Department of Astrophysics, IMAPP, Radboud University Nijmegen, PO Box 9010, NL-6500 GL, Nijmegen, The Netherlands}
\altaffiltext{10} {School of Physics and Astronomy, University of Birmingham, UK}

\altaffiltext{11}{Astronomical Institute Anton Pannekoek, University of Amsterdam, The Netherlands}
\altaffiltext{12} {Max Planck Institute for Solar System Research, Justus-von-Liebig-Weg 3
37077 G\"ottingen, Germany}
\altaffiltext{13}{LESIA, PSL Research University, CNRS, Universit\`e Pierre et Marie Curie, Universit\`e Denis Diderot, Observatoire de Paris, Meudon Cedex, France}
%%%%%%%%%%%%%%%%%%%%%%%%%%%%%%%%%%%%%%%%%%%%%%%%%%%%%%%%%%%%%%%%%%%%%%%%%%%
\begin{abstract}

 In this paper we study the dynamics of the stellar interior of
the early red-giant star KIC~4448777 by asteroseismic inversion of
14 splittings  of the dipole mixed modes obtained from {\it Kepler} observations.
 In order to overcome  the complexity of the oscillation pattern typical of 
red-giant stars, we present a procedure which involves a combination of different methods to extract the rotational splittings from
the power spectrum.

We find not only that the
 core rotates faster than the surface, confirming previous inversion results generated for other red giants \citep{deheuvels2012,deheuvels2014}, but we also
  estimate the variation of the angular velocity within the helium core with a spatial resolution of 
$\Delta r=0.001R$ and verify the hypothesis of a sharp discontinuity in the inner stellar rotation \citep{deheuvels2014}.
 The results show that the entire core rotates rigidly with an angular velocity  of about $\langle\Omega_c/2\pi\rangle=748\pm18$~nHz and 
 provide evidence for an angular velocity  decrease
 through a region between the helium core and part of the hydrogen burning shell; however we
   do not succeed to characterize the rotational slope, due to the intrinsic limits of the applied techniques.
 The angular velocity, from the edge of the core and through the hydrogen burning shell, appears to decrease with
increasing distance from the center,
 reaching an average value  in the convective envelope of $\langle\Omega_s/2\pi\rangle=68\pm22$~nHz. Hence,  the core in KIC~4448777 is rotating from a minimum of 8 to a maximum of 17 times faster than the envelope.

 We conclude that a set of data which includes only dipolar modes is sufficient to infer quite accurately the rotation of a red giant not only in the dense core but also,  with a lower level of confidence,
 in part of the radiative region and in the convective envelope.

 \end{abstract}
%%%%%%%%%%%%%%%%%%%%%%%%%%%%%%%%%%%%%%%%%%%%%%%%%%%%%%%%%%%%%%%%%%%%%%%%%%%
%%%%%%%%%%%%%%%%%%%%%%%%%%%%%%%%%%%%%%%%%%%%%%%%%%%%%%%%%%%%%%%%%%%%%%%%%%%
\section{Introduction}
Stellar rotation is one of the fundamental processes governing stellar structure and evolution.
The internal structure of a star at a given phase of its life is strongly affected by the angular momentum transport history. Investigating the internal rotational profile of a star and reconstructing its evolution with time become crucial in achieving basic constraints on the angular momentum transport mechanisms acting in the stellar interior
during different phases of its evolution. In particular, physical processes that affect rotation and in turn are affected by rotation, such as convection, turbulent viscosity, meridional circulation, mixing of elements, internal gravity waves, dynamos and magnetism, are at present not well understood and modeled with limited success \citep[e.g.,][]{marques2013, cantiello}.

Until fairly recently, rotation inside stars has been a largely unexplored field of research from an observational point of view. Over the past two decades helioseismology changed this scenario, making it possible to measure the rotation profile in the Sun's interior through the measurement of the splittings of the oscillation frequencies, revealing a picture of the solar internal dynamics very different from previous theoretical predictions \citep[see e.g.,][]{elsworth1995,schou1998,thompson2003}.  
 Contrary to what one could expect from the angular momentum conservation theory,
predicting a Sun
 with a core rotating much faster than the surface 
 when only meridional circulation and classical hydrodynamic instabilities are invoked \citep[e.g.,][]{chaboyer},
  helioseismology shows an almost uniform rotation in the radiative interior  and an angular velocity monotonically decreasing from the equator to high latitudes in the convective envelope.
 This strongly supports the idea that several powerful processes act to redistribute angular momentum in the interior, like for example magnetic torquing \citep[e.g.,][]{brun}
and internal gravity waves \citep[e.g.,][]{talon2005}.

The rotation breaks the spherical symmetry of the stellar structure and splits
the frequency of each oscillation mode of harmonic degree $l$ into 
$2l+1$ components which appear as a multiplet in the power spectrum. Multiplets with a fixed radial order $n$ and harmonic degree $l$ are said to exhibit a frequency ``splitting'' defined by:
\begin{equation}
\delta \nu_{n,l,m}=\nu_{n,l,m}-\nu_{n,l,0}\; ,
\label{Eq.1}
\end{equation}
somewhat analogous to the Zeeman effect on the degenerate energy levels of an atom, where $m$ is known as the azimuthal order.
 Under the hypothesis that the rotation of the star is sufficiently slow, so that effects of the centrifugal force can be neglected, the frequency separation between components of the multiplet is directly related to the angular velocity \citep{cow}.

In recent years spectacular asteroseismic results on data provided by the space missions CoRoT \citep{baglin2006} and {\it Kepler} \citep{borucki2010} have revolutionized the field. In particular, the {\it Kepler} satellite has provided photometric time series of unprecedented quality, cadence and duration, supplying the basic conditions for studying the internal rotational profile and its temporal evolution in a large sample of stars, characterized by a wide range of masses and evolutionary stages.

In this context the detection of solar-like pulsations - as in the Sun excited by turbulent convection -  in thousands of red giants, from the bottom to the tip of the red-giants branch  \citep[see, e.g.,][]{mosser2013b, Stello13} and to the AGB \citep{corsaro2013} appears particularly exciting. Red-giant stars are ideal asteroseismic targets for many reasons.
Compared to the main-sequence stars, solar-like oscillations in red giants are easier to detect due to their higher pulsation amplitudes \citep{mosser2013a}.
%Furthermore - and even more powerful for asteroseismology - unlike the  Fourier spectra of solar-like main sequence stars which show the characteristic frequency comb-like pattern of high-order acoustic modes, those of red giants exhibit a dense spectrum of non-radial modes with mixed g-p character \citep[see, e.g.,][]{beck2011}.
What is more important,
red-giant frequency spectra
reveal mixed modes \citep[see, e.g.,][]{beck2011},
which probe not only the outer layers, where they behave like acoustic modes, but also the deep radiative interior, where they
propagate as gravity waves.
Both the gravity
and the acoustic-wave propagation zones contribute, in various proportions, to the
formation of mixed modes.
The greatest contribution from the acoustic zone occurs for modes with frequency near
the resonant frequency of
the acoustic cavity. 
%The frequency difference between these more acoustic modes is almost
%the same as between
%consecutive radial modes. 
 Mode inertias attain
local minima at these frequencies.

Moreover, the red-giant phase represents a crucial step in the stellar angular momentum distribution history \citep{ceillier2013,marques2013}. When a star evolves off the relatively long and stable main sequence, its rotation starts evolving differently in the inner and outer parts causing the formation of a sharp rotation gradient in the intermediate regions where hydrogen is still burning:  assuming that the angular momentum is locally conserved,
 the contraction of the core causes its rotation to speed up
  in a relatively short timescale, while the outer layers slow down due to their expansion. The accurate determination of the rotational profiles in subgiants and red giants provides information on the angular momentum transport mechanism potentially leading to significant improvements in the
modeling of stellar structure and evolution.

Recently, results  based on measurements of the rotational splittings of dipole mixed modes have been reported in the literature \citep[e.g.][]{beck2012, beck2014, mosser2012c, deheuvels2012, deheuvels2014}. 
 \citet{beck2012}, based on high precision measurements of rotational splittings provided by {\it Kepler},
 found that
the core in the red-giant stars
is rotating faster than the upper layers. These results were confirmed by applying inversion techniques to rotational splittings by \citet{deheuvels2012, deheuvels2014}.  Asteroseismology of large sample of stars \citep{mosser2012c, deheuvels2014}  allowed to clarify that the mean core rotation 
 significantly slows down as stars ascend the red-giant branch.

Several theoretical investigations have explored the consequences of these recent results on internal angular momentum transport inside solar-like oscillating stars along their evolution \citep[e.g.,][]{ceillier2013, marques2013, tayar2013, cantiello}.
These results show that the internal rotation rates, predicted by current theoretical models of subgiants and red giants, are at least 10 times higher compared to observations, suggesting the need to investigate more efficient mechanisms of angular-momentum transport acting on the appropriate timescales during these phases of stellar evolution.

In this paper we analyze more than two years of {\it Kepler} observations of the red-giant star KIC~4448777 and identify 14 rotational splittings of mixed modes in order to characterize its internal rotational profile
using different inversion techniques  at first applied successfully to helioseismic data \citep[e.g.,][]{thompson96, schou1998, paterno1996, dimauro1998} and recently
to data of more evolved stars \citep{deheuvels2012, deheuvels2014}.
 
The paper is organized
as follows: Section 2 reports the results of the spectroscopic analysis of the star aimed at the determination of its atmospheric parameters. Section 3 describes the method adopted to analyze the oscillation spectrum and identify the mode frequencies and the related splittings. Section 4 provides the basic formalism for performing the inversion, starting from the observed splittings and models of the star.  Section 5 describes the evolutionary models constructed to best fit the atmospheric and asteroseismic constraints. Section 6 presents the details of the asteroseismic inversion carried out to infer the rotational profile of the star.  In Section 7 we test the inversion techniques for the case of red giants and the capability  of detecting  the presence of rotational gradient  in the deep interior of the star. In Section 8 the results obtained by the inversion techniques are compared with those obtained by other methods.
Section 9 summarizes the results and draws the conclusions.

\section{Spectroscopic analysis}
\label{Sec:spec}

In order to properly characterize the star, six spectra of 1800 seconds integration time each were obtained with the HERMES spectrograph \citep{raskin2011}, mounted on the 1.2-m MERCATOR telescope at La Palma. This highly efficient \'echelle spectrometer has a spectral resolution of R=86000, covering a spectral range from 380 to 900 nm. The raw spectra were reduced with the instrument specific pipeline and then averaged to a master spectrum. The signal-to-noise ratio was around 135 in the range from 500 to 550 nm.

The atmospheric parameter determination was based upon Fe I and Fe II lines which are abundantly present in red-giant spectra. We used the local thermal equilibrium (LTE) Kurucz-Castelli atmosphere models \citep{castelli} combined with the LTE abundance calculation routine MOOG (version August 2010) by C. Sneden. Fe lines were identified using VALD line lists \citep{kupka}. For a detailed description of the different steps needed for atmospheric parameter determination, see, e.g., \citet{desmedt}.
We selected Fe lines in the highest signal-to-noise region of the master spectrum in the wavelength range between 500 and 680 nm. The equivalent width (EW) was calculated using direct integration and the abundance of each line was then computed by an iterative process where theoretically calculated EWs were matched to observed EWs. Due to the high metallicity, the spectrum of KIC~4448777 displays many blended lines. To avoid these blended lines in our selected Fe line lists, we first calculated the theoretical EW of all available Fe I and Fe II lines in the wavelength range between 500 and 680 nm. The theoretical EWs were then compared to the observed EWs to detect any possible blends. The atmospheric stellar parameters derived by the spectroscopic analysis are reported in Table ~\ref{tab:parameters} and are based upon the results from 46 Fe I and 32 Fe II lines.

\begin{deluxetable}{ccccc}
 \tablewidth{0pc}
 \tablecolumns{5}
 \tablecaption{KIC~4448777 atmospheric parameters}
\tablehead{\colhead{M$_V$ } & \colhead{$T_{\mathrm{eff}}$ (K)} & \colhead{$\log g$ (dex)}  &\colhead{$[ \mathrm{Fe/H}]$} & \colhead{$v\sin i$ (km s$^{-1}$)}}
\startdata
 $11.56$ \tablenotemark{1} & $4750  \pm 250$  & $3.5 \pm 0.5$ & $0.23 \pm 0.12$ & $< 5$ \\
 \enddata
 \tablenotetext{1}{{\it Kepler} catalogue}
\label{tab:parameters}
 \end{deluxetable}

We have also explored the possibility to derive the surface rotation rate by following the method by \citet{Garcia2014}, but we have not
found any signatures of spot modulation 
as evidence for an on-going magnetic field.

\section{Time series analysis and Fourier spectrum}
\label{sec:osc}
For the asteroseismic analysis we have used near-continuous photometric time
series obtained by {\it Kepler} in long-cadence mode (time sampling of
29.4 min).  This light curve spans about 25 months corresponding to
observing quarters Q0-9, providing a formal frequency resolution of 15~nHz.
We used the so-called PDC-SAP (pre data conditioning - simple aperture photometry) light curve \citep{Jenkins10} corrected for
instrumental trends and discontinuities as described by \citet{Garcia11}.

\begin{figure}[!htb]
\centering
\includegraphics[width=12cm]{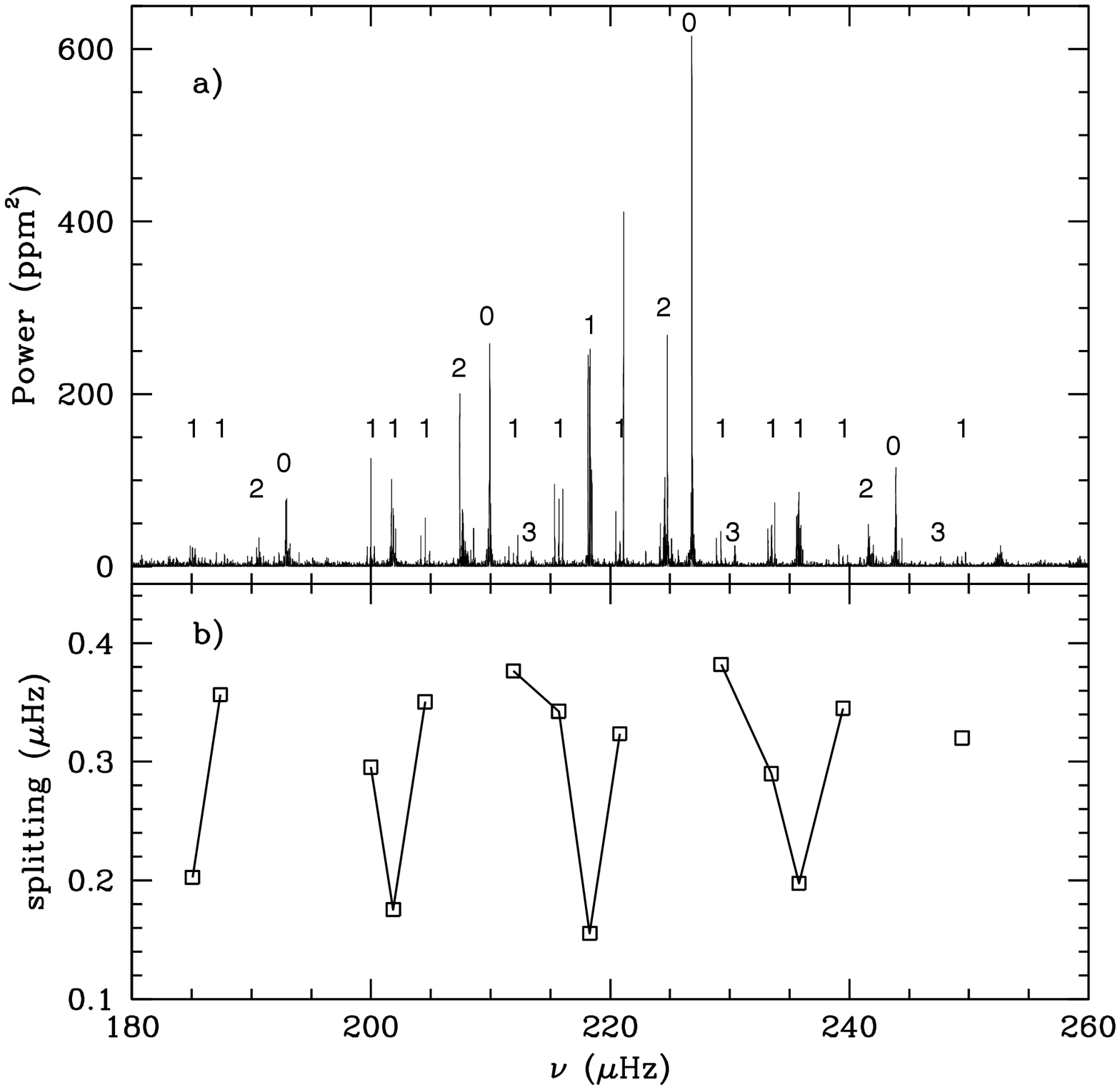}
\vspace{-3cm}
\caption{In panel a) the observed frequency spectrum of
  KIC~4448777 is shown. The harmonic degrees of the modes ($l=0,1,2,3$)
  are indicated.  Multiplets due to rotation are visible for $l=1$.
  Panel b) shows observed rotational splittings for $l=1$ modes.}
\label{fig:spectrum}
\end{figure}

The power spectrum of the light curve shows a 
clear power excess in the range $(180-260)\, \mu$Hz
(Fig. \ref{fig:spectrum}) due to radial modes, with the comb-like pattern typical of the solar-like p-mode
oscillations, and non-radial modes, particularly those of spherical degree $l=1$,
modulated by the mixing with g modes. 

The initial analysis of the spectrum was done using the pipeline described
in \citet{Huber09}.  By this method we determined the frequency at maximum
oscillation power $\nu_{\rm max} = (219.75 \pm 1.23)\, \mu$Hz and the
so-called large frequency separation between modes with the same harmonic degree
$\Delta\nu = (16.96 \pm 0.03)\, \mu$Hz.  The quoted uncertainties have been derived from analyzing 500 spectra generated by randomly perturbing the original
spectrum, according to \citet{huber2012}. For the purposes of this paper
it has been necessary, also, to
identify the individual modes and measure their frequencies.  This process,
known as "peak bagging", is notoriously
difficult to perform for red giants because of their complex frequency
spectra, with modes of very different characteristics within narrow, sometimes
even overlapping, frequency ranges.  Mixed modes of different inertia have very different damping times and hence also different profiles in the frequency spectrum.  Here  we therefore used a combination
of known methods tailored to our particular case, although we should mention that
 an automatic
"one-fits-all" approach has been recently developed by \citet{corsaro}.

Four independent groups (simply called ’fitters’) performed the fitting of the modes by using  slightly different approaches: 
 \begin{enumerate}
 \item The first team smoothed the power spectrum of the star to account
 for the intrinsic damping and reexcitation of the modes.
They located the modes by two separate steps using a different level of smoothing in each.
 First, they heavily smoothed the power spectrum 
 (by 13 independent bins in frequency) to detect the most damped modes including radial, quadrupole and dipole modes with lower inertia, and then they smoothed less (by 7 bins)  to  identify the dipole modes with higher inertia. 
   In both cases the peaks were selected and associated with modes only if they were significant at the 99\% level,  setting the threshold according to the statistics for smoothed spectra, which takes into account the level of smoothing applied and the frequency range over which the modes have been searched  \citep[see e.g.,][]{Chaplin2002}. 
 In addition, the `toy model' of \citet{Stello12} was used to locate a few extra dipole modes.
 % that had fallen just below the nominal 99\% significance level. 
 This fitting was performed using the MCMC Bayesian method by 
 \citet{HandbergCampante11};

 \item The second team extracted the frequencies of individual modes as the centroid of the unsmoothed power spectral density (PSD)  in narrow predefined windows, checked for consistency by fitting Lorentzian profiles to a number of modes  \citep{beck2013};
   
 \item The third team modelled the power spectrum with a sum of many Lorentzians,
 performed a global fit using a maximum likelihood estimator (MLE), and calculated the 
 formal uncertainties from the inverse Hessian
 matrix  \citep{mathur2013};
 
 \item The fourth team derived proxies of the oscillation frequencies from a global
 fit based on global seismic parameters; all peaks with a height-to-background ratio larger than 8 were selected; radial modes and quadrupole modes were estimated from the fit of the large separation provided by the second-order asymptotic expansion for pressure modes \citep{mosser2011}; the dipole modes
 were obtained with the asymptotic expansion for mixed modes \citet{mosser2012b} with rotational splittings derived with the method of \citet{goupil2013}.
 
 \end{enumerate}

  The final set of 58 individual mode frequencies, including the multiplets due to rotation for the $l=1$ modes, consists only of those frequencies detected at least by two of the fitters.

In order to obtain statistically consistent uncertainties for the mode
frequencies we used the Bayesian Markov-Chain Monte Carlo (MCMC) fitting algorithm
by \citet{HandbergCampante11} for the peak bagging.  The algorithm allows
simultaneous fitting of the stellar granulation signal \citep{mathur2011} and all oscillation
modes, each represented by a Lorentzian profile.
However, due to the complexity of the frequency spectrum of KIC~4448777, we
were not able to fit all modes simultaneously using a single method.  In
particular, the mixed modes with very high inertia and hence
very long mode lifetimes have essentially undersampled frequency profiles
in the spectrum.  Fitting Lorentzian profiles to these modes is therefore
unsuitable, and can easily lead the fitting algorithm astray. 

 We therefore
treated the radial and quadrupole modes separately from the dipole modes.
The radial and quadrupole modes were fitted as Lorentzian profiles  using the
MCMC approach.  In this analysis we ignored any mixing of quadrupole modes
and hence fit only one quadrupole mode per radial order.

For the dipole modes, which we were not able to fit as part of the MCMC method for the reasons explained above, we decided to adopt the initial frequencies, found from the smoothed spectrum by Team 1, as the final frequencies and the scatter among the four fitters as a proxy for uncertainties.
%, since  the data did not allow the fitters to obtain robust uncertainties for the modes in question.
 As a sanity check for this approach we compared the uncertainties obtained by the MCMC fitting procedure of the radial modes with the scatter between different fitters of the same modes. We found that on average the ‘fitter scatter’ is within 17\% of the MCMC uncertainty, and all fitter scatter values are within a factor of two of the MCMC-derived uncertainty.

In the above analysis each dipole mode was detected separately,
independently of the azimuthal order $m$. 
As in \citet{beck2014},
 we noticed that the components with $m=\pm1$
of a given triplet are not equally spaced from the
central $m=0$ mode.  In the framework of the perturbation theory, the splitting asymmetries correspond to second-order effects in the oscillation frequency
that mainly
account for the distortion caused by the centrifugal force.
Here, the asymmetry
is smaller in size 
and reasonably negligible, in first approximation, when compared to the rotational splitting itself, ranging from 0.3\% to - at most - 12\% (with a mean value of 6\%) of the respective rotational splittings, with values comparable with the frequency uncertainties.
In order to remove second order perturbation effects, here we used the generalized rotational splitting expression \citep[e.g.,][]{goupil2009}:
 \begin{equation}
\delta \nu_{n,l}=\frac{\nu_{n,l,m}-\nu_{n,l,-m}}{2m}.
\end{equation}
The relative uncertainties have been calculated according to the general propagation formula for indirect measurements.
Detailed investigation on the physical meaning of rotational splitting asymmetries  and the possibility to derive from them more stringent constraints on the internal rotational profile of oscillating stars have been the subject of different papers \citep[cf.,][]{suarez}, but it is beyond the aim of the present work.

 Table~\ref{tab:frequencies} lists the final set of frequencies together
with their uncertainties, corresponding to the values obtained by the MCMC fitting procedure for radial and quadrupole modes and to the scatter in the results from the four fitters for dipole modes,  their spherical degree and the rotational splittings for 14 dipole modes.

To measure the inclination of
the star we used the above MCMC peak bagging algorithm, restricting
the fit to the strongest dipole modes of $m=0$ and imposing equal spacing of
the $m=\pm1$ components. This calculation provided an inclination of
$i=32.6^ {+5.0}_{-4.3}\deg$.

As in \citet{beck2012}, the observed rotational splittings are not constant for consecutive dipole modes (see
Fig. \ref{fig:spectrum}b showing rotational splittings for the $l=1$ modes).
Splittings are larger for modes with a higher inertia
which predominately probe the inner radiative interior. This shows that the
deep interior of the star is rotating faster than the outer layers.

The identification of several dipole 
modes and the use of the method by
\citet{mosser2012b}, based on the asymptotic relation, allowed us to estimate the asymptotic period spacing $\Delta \Pi_1 = 89.87 \pm 0.07\,$s, which places this star on the low luminosity red-giant branch in agreement with the evolutionary phase predicted by
the value of the observed $\Delta\nu$ 
 \citep{bedding2011,mosser2012b,Stello13}.

A first
estimate of the asteroseismic stellar mass and radius can be obtained
from the observed $\Delta\nu$ and $\nu_{\mathrm{max}}$ together with the value of $T_{\mathrm{eff}}$
 \citep{KjeldsenBedding95,kallinger2010, belkacem2011, miglio2012}.
 In particular, by using the scaling
relation calibrated on
 solar values, we obtain $M_{ast}/{\mathrm M}_{\odot} = 1.12 \pm 0.09$
 and $R_{ast}/{\mathrm R}_{\odot} = 4.13 \pm 0.11$.  
By using the scaling relation by \citet{mosser2013a}, calibrated on a large sample of observed solar-like stars, we obtain
 $M_{ast}/{\mathrm M}_{\odot} = 1.02 \pm 0.05$
  and $R_{ast}/{\mathrm R}_{\odot} = 3.97 \pm 0.06$.
 From the above values we can determine the asteroseismic surface gravity be $\log g_{ast}=3.25\pm0.03$~dex. This value is in good agreement with that determined by the spectroscopic analysis (see Table ~\ref{tab:parameters}).

\begin{deluxetable}{cccccc}
\tabletypesize{\scriptsize}
\tablecolumns{6}
\tablewidth{0pt}
\tablecaption{Measured central frequencies ($m=0$) and rotational splittings for KIC~4448777.}
\tablehead{\colhead{$l$} & \colhead{$\nu_{n,l} (\mu{\mathrm Hz})$}  & \colhead{$\delta\nu_{n,l} (\mu {\mathrm Hz})$} & %\hspace*{0.7cm}
\colhead{\hspace*{0.7cm}$l$} & \colhead{$\nu_{n,l} (\mu{\mathrm Hz})$}  & \colhead{$\delta\nu_{n,l} (\mu {\mathrm Hz})$}}
\startdata
 0  & 159.842 $\pm$ 0.014  & n.a.\tablenotemark{1} & \hspace*{0.7cm} 2  &   174.005 $\pm$ 0.043 &... \\
 0  & 176.277 $\pm$ 0.018  & n.a.                 & \hspace*{0.7cm} 2  &  190.623 $\pm$ 0.034  & ... \\
 0  & 192.907 $\pm$ 0.016  & n.a.                 & \hspace*{0.7cm} 2  &  207.551  $\pm$ 0.026 & ...\\
 0  & 209.929 $\pm$ 0.014   & n.a.                &\hspace*{0.7cm} 2  &  224.646  $\pm$ 0.011 & ...\\
 0  & 226.831 $\pm$ 0.014   & n.a.                &\hspace*{0.7cm} 2  &   241.630  $\pm$ 0.022 & ...\\
 0  & 243.879 $\pm$ 0.013   & n.a.                &\hspace*{0.7cm} 3 & 213.443 $\pm$ 0.015 & ...\\
 0  & 261.215 $\pm$ 0.034   & n.a.               &\hspace*{0.7cm} 3 & 230.423 $\pm$ 0.011 & ...\\
 1  & 167.061 $\pm$ 0.011   & ...                &\hspace*{0.7cm} 3 & 247.600 $\pm$ 0.017 & ...\\
 1 &    185.069 $\pm$ 0.011\tablenotemark{2} &  0.2025 $\pm$ 0.0078 &                &  &\\
 1 &   187.402  $\pm$ 0.011\tablenotemark{2}& 0.3565 $\pm$ 0.0078&                  & &\\
 1 &  199.986  $\pm$ 0.011\tablenotemark{2}  & 0.2955 $\pm$ 0.0078&                  & & \\
 1 &    201.864  $\pm$ 0.011\tablenotemark{2}  & 0.1755 $\pm$ 0.0078& & &  \\
 1 &  204.528    $\pm$ 0.011\tablenotemark{2} & 0.3505 $\pm$ 0.0078 & & & \\
1  &  208.571    $\pm$ 0.011\tablenotemark{2} & ...                 & & &  \\
 1 &    211.913  $\pm$ 0.011\tablenotemark{2} & 0.3765 $\pm$ 0.0078 & & & \\
 1 &   215.699   $\pm$ 0.015\tablenotemark{2} & 0.3425 $\pm$ 0.0078 & & &\\
 1  &  218.299   $\pm$ 0.017\tablenotemark{2} & 0.1555 $\pm$ 0.0078 & & &\\
 1  &  220.814   $\pm$ 0.018\tablenotemark{2} & 0.3235 $\pm$ 0.0078 & & &\\
 1 &    229.276   $\pm$ 0.011\tablenotemark{2} & 0.3820 $\pm$ 0.0160 & & &\\
 1 &    233.481 $\pm$ 0.011\tablenotemark{2}   & 0.2900 $\pm$ 0.0078 & & &\\
 1 &    235.783   $\pm$ 0.011\tablenotemark{2} & 0.1975 $\pm$ 0.0140 & & &\\
 1 &    239.463   $\pm$ 0.011\tablenotemark{2} & 0.3450 $\pm$ 0.0160 & & &\\
 1 &   244.385    $\pm$ 0.011\tablenotemark{2} & ...                 & & & \\
 1 &    249.417    $\pm$ 0.011\tablenotemark{2}  & 0.3200 $\pm$ 0.0210& & &\\
 1 &   252.377    $\pm$ 0.021\tablenotemark{2} &  ... & & &\\
 %1 &   252.377    $\pm$ 0.021 & 0.2840 $\pm$ 0.0050 & & &\\
 1 &   252.661    $\pm$ 0.016\tablenotemark{2} & ...& & &\\
\enddata
      \tablenotetext{1}{n.a.= not applicable}
      \tablenotetext{2}{Uncertainties for the $l=1$ modes correspond to the scatter in the results from the four fitter (see text)}
\label{tab:frequencies}
\end{deluxetable}

%%%%%%%%%%%%%%%%%%%%%%%%%%%%%%%%%%%%%%%%%%%%%%%%%%%%%%%%%%%%%%%%%%%%%%%%%%%

\section{Asteroseismic inversion}
The asteroseismic inversion is a powerful tool which allows to
estimate the physical properties of stars,
by solving integral equations expressed in terms of the experimental data.

Previous experience acquired in helioseismology
on inverting solar data represents a useful background for asteroseismic inversion.
Earlier attempts in generalizing the standard helioseismic differential methods to find the structure differences between the observed star and a model have been applied to artificial data with encouraging results by \citet{GK93}, \citet{roxburgh98}, \citet{Ber01}.  More recently, \citet{dim04}
was able to infer the internal structure of Procyon~A below $0.3 R$ 
 by inversion of real data comprising 55 low-degree p-mode frequencies
 observed in the star. 
A general conclusion from these previous investigations is that the success of the inversion depends strongly on the number of observed frequencies and the accuracy with which the model represents the star.

Stellar inversions to infer the internal rotational profiles of stars
were firstly applied to
artificial data of moderately rotating stars such as 
 $\delta$-Scuti stars \citep{goupil96}
and white dwarfs \citep{kaw99}.
Rotational inversion of simulated data of solar-like stars was studied by 
\citet{lochard05} for the case of a subgiant model representative of $\eta$ Boo. They showed that mixed modes can improve  the inversion results on the internal rotation of the star, while data limited to pure $l=1,2$ p modes are not sufficient to provide reliable solutions.
Indeed, striking results on rotation have been obtained by \citet{deheuvels2012} who performed a detailed modeling of the red-giant star KIC~7341231, located at the bottom of the red giant branch. They performed an inversion of the internal stellar rotation profile based on observed rotational splittings of 18 mixed modes. They found that the core is rotating at least five times faster than the envelope. More recently \citet{deheuvels2014} applied their techniques to six subgiants and low-luminosity red giants.

The internal rotation of KIC~4448777 can be quantified by inverting the following equation
 \citep{gough1981}, obtained by applying a standard perturbation theory to the eigenfrequencies, in the hypothesis of slow rotation:
\begin{equation}
\delta\nu_{n,l} = \int_{0}^{R} {\cal K}_{n,l}(r) \frac{\Omega(r)}{2 \pi}\, dr +\epsilon_{n,l}\; ,
\label{eq:rot}
\end{equation}
where  $\delta\nu_{n,l}$ is the adopted set of splittings, $\Omega(r)$ is the internal rotation assumed to be a function of only the radial coordinate, $\epsilon_{n,l}$ are the uncertainties in the measured splittings and ${\cal K}_{n,l}(r)$ is
 the mode kernel functions calculated on the unperturbed eigenfunctions for the modes $(n,l)$ of {\it the best model} of the star:
\begin{equation}
{\cal K}_{n,l}(r)=\frac{1}{I}\left[\xi_{r}^{2}+l(l+1)\xi_{h}^{2}-2\xi_{r}\xi_{h}-\xi_{h}^{2}\right]\rho r^{2}\; ,
\label{ker}
\end{equation}
where $\xi_{r}$ and $\xi_{h}$ are the radial and horizontal components of the displacement vector respectively, $\rho$ is the density and $R$ is the photospheric stellar radius, while the inertia is given by:
\begin{equation}
I=\int_{0}^{R}\left[\xi_{r}^{2}+l(l+1)\xi_{h}^{2}\right]\rho r^2 dr\; .
\label{inertia}
\end{equation}

The properties of the inversion
 depend both on the mode selection $i\equiv(n,l)$
and on the observational uncertainties $\epsilon_{i}$
which characterize {\it the mode set}
$i= 1, \ldots, N$ to be inverted.

The main difficulty in solving Eq. \ref{eq:rot} for $\Omega(r)$ arises from the fact that
 the inversion is an
ill-posed problem: the observed splittings constitute a
finite and quite small set of data and 
 the uncertainties in the observations prevent the solution from being determined
with certainty.
Thus, an appropriate choice for a
 suitable inversion technique is
the first important step during an asteroseismic inverse analysis.

\subsection{Inversion procedure}

There are two important classes of methods for obtaining estimates of $\Omega(r)$ from Eq. 
\ref{eq:rot}: 
the optimally localized averaging (OLA) method, based on the original idea of
\citet{backus1970}, and 
the regularized least-squares (RLS) fitting method \citep{ph62, ti63}. Both methods give linear estimates of
the function $\Omega(r)$ with results generally in 
 agreement, as was demonstrated by
\citet{CD90, sekii97, deheuvels2012}.

 Here we study and apply the OLA method
and its variant form,
 which allows us to estimate a
localized weighted average of angular velocity $\bar{\Omega}(r_{0})$ at selected target radii $\{r_{0}\}$ by means of a linear combination of all the data:
\begin{equation}
\frac{\bar{\Omega}(r_{0})}{2\pi}=\sum_{i=1}^{N}c_{i}(r_{0})\delta\nu_{i}= \int_{0}^{R} {K}(r_0,r) \frac{\Omega(r)}{2\pi}dr\; ,
\label{backus}
\end{equation}
where $c_{i}(r_{0})$ are the inversion coefficients and 
\begin{equation}
K(r_{0},r)=\sum_{i=1}^{N} c_{i}(r_{0}){\cal K}_{i}(r)
\end{equation}
 are the averaging kernels.
 Here we adapted the code, developed for solar rotation in
\citet{paterno1996}, to be applied to any evolutionary phase.

Because of the ill-conditioned nature of the inversion problem, 
it is necessary to introduce a regularization procedure.
By varying a trade-off parameter $\mu$,
%(\mu Hz)^{-2} or s^2
we look for the coefficients $c_{i}(r_{0})$ that minimize the propagation of the uncertainties
and the spread of the kernels:
\begin{equation}
\int_{0}^{R}J(r_0,r)K(r_{0},r)
^2 dr+\frac{\mu}{ \mu_0} \sum_{i=1}^{N}\epsilon_{i}^2
c_{i}^2(r_{0})\; ,
\label{ls}
\end{equation}
where 
\begin{equation}
\mu_0=\frac{1}{N}\sum_{i=1}^{N}\epsilon_{i}^2
\end{equation}
assuming that
\begin{equation}
\int_{0}^{R}
K(r_{0},r) dr=1 \; .
\end{equation}
$J(r_0,r)$ is a weight function,
small near $r_0$ and large elsewhere,  which has been assumed to be:
\begin{equation}
J(r_0,r)=12(r-r_0)^2/R\;,
\end{equation}
designed to build averaging kernels as close as possible to a Dirac function centered
in $r_0$.
The minimization of Eq.~\ref{ls} is equivalent to solve a set of $N$ linear equations for $c_i$.
The uncertainties of the solutions are the standard deviations calculated in the following way:
\begin{equation}
\sigma\left[\frac{\bar{\Omega}(r_0)}{2\pi}\right]=\left[\sum_{i=1}^N c_i^2(r_0)\epsilon_i^2\right]^{1/2}\; .
\label{deltaf}
\end{equation}
%while the radial spatial resolution is assumed to be the half width at half maximum of the averaging kernels.
The center of mass of the averaging kernels is:
%define the location of the solutions:
\begin{equation}
\bar{r}(r_{0})=\int_0^R {r}K(r_{0},r)dr.
\label{cgravity}
\end{equation}

We also considered the method in the variant form, as described in 
\citet{pijpers1992}, known as SOLA (Subtractive Optimally Localized Averaging),
 making attempts to fit the averaging kernel to a Gaussian function $G(r_{0},r)$ of an appropriate width, centered at the target radius \citep{dimauro1998}.
The two parameters, the width of the Gaussian target function
and the trade-off parameter, are tuned to find an acceptable matching of the averaging kernel
to its target function and also to ensure an acceptable small error on the result
from the propagation of the measurement errors.
Therefore, the coefficients are determined by minimizing the following:
\begin{equation}
 \int_{0}^{R}
R\left[K(r_{0},r)-G(r_{0},r
)\right]^2 dr+\frac{\mu}{\mu_0}\sum_{i=1}^{N}\epsilon_i^2
c_{i}^2(r_{0}), 
\label{sola}
\end{equation}
 where
\begin{equation}
G(r_{0},r
)=\frac{1}{\sqrt{2\pi\sigma^2}}\exp^{-|r-r_0|^2/2\sigma^2}
\end{equation}
and  $\sigma$ is chosen to fix the width of the Gaussian function.
% and
%the full width at half maximum of the Gaussian function is one of the fixed parameter
%$\Delta=2\sqrt{\ln2}\sigma$.

\section{Evolutionary models of KIC~4448777}

 We first need to construct a {\it
  best fitting} model of the star that satisfies all the observational
constraints in order to
quantify the internal rotation and to understand the relation between the observed rotational splittings and how sensitive each mode is to the different regions
of that model.  
The theoretical structure models have been calculated by using the ASTEC evolution code \citep{chris2008a}, spanning the parameter space 
given in Table \ref{tab:parameters} and following the
procedure described in \citet{dimauro2011}.

The input physics for the evolution calculations included the OPAL 2005 equation of state 
\citep{OPAL}, OPAL opacities % (Iglesias \& Rogers 1996),
\citep{Igl96}, and the NACRE nuclear reaction rates \citep{NACRE}. Convection was treated according to the
mixing-length formalism (MLT) \citep{bohm} and defined through the parameter $\alpha=\ell/H_p$, where $H_p$ is the pressure scale height and $\alpha$ is varied from $1.6$ to $1.8$.
The initial heavy-element mass fraction $Z_i$
has been calculated from the iron abundance given in Table \ref{tab:parameters} using the relation [Fe/H]$=\log(Z/X)-\log(Z/X)_{\odot}$, where $(Z/X)$ is the value at
the stellar surface and the solar value was taken to be $(Z/X)_{\odot}=0.0245$ \citep{GN93}. Thus, we used
 $Z/X=0.04\pm0.01$ in the modeling.

The resulting evolutionary tracks are characterized by the input stellar mass $M$, the initial chemical
composition and a mixing-length parameter. 
For the models with values
of $T_\mathrm{eff}$ and $\log g$ consistent with the spectroscopic observed values, we calculated the 
adiabatic oscillation frequencies using the ADIPLS code \citep{chris2008b}.
We applied 
the surface effect correction following the approach proposed by \citet{kjeldsen2008} and using the prescription of \citet{brandao11}, which takes into account that modes with high inertia suffer a smaller surface effect than do p modes.
 The correction applied to all calculated frequencies is then of the form:
\begin{equation}
\nu^{mod}_{n,l}=\nu_{n,l}+a\frac{1}{Q_{n,l}}\left(\frac{\nu_{n,l}}{\nu_0}\right)^b
\end{equation}
where $\nu^{mod}_{n,l}$ are the corrected frequencies, $Q_{n,l}$ is the inertia of the given mode normalized by the inertia of a radial mode of the same frequency,  obtained by interpolation, $\nu_{n,l}$
 are the best-model frequencies,
   $\nu_0$ is a constant frequency, usually chosen to be the frequency at maximum
oscillation power, $a$ is the amplitude of the correction at $\nu_0$ and $b$ is the exponent assumed to be $4.90$ as the one calculated for the solar frequencies by \citet{kjeldsen2008}. 

The results of the fits between the observed star and the models were evaluated
according to the total $\chi^2$
 between the observed $\nu^{obs}_i$ and calculated $\nu^{mod}_i$ values of the individual
oscillation frequencies as:
\begin{equation}
\chi^2=\frac{1}{N}\sum_1^N\left(\frac{\nu^{obs}_i-\nu^{mod}_{i}}{\epsilon_i}\right)^2,
\end{equation}
where $\epsilon_i$ are the uncertainties on the observed frequencies.
\begin{deluxetable}{lccc}
\tabletypesize{\scriptsize}
\tablewidth{0pc}
 \tablecolumns{6}
 \tablecaption{Main 
 parameters for KIC~4448777 and for the best fitting models.}
 \tablehead{\colhead{} & \colhead{KIC~4448777}&\colhead{ Model 1}& \colhead{Model 2}}
\startdata
 $M/{\mathrm M}_{\odot}$ &$1.02\pm0.05\tablenotemark{a}$ & 1.02 &  1.13\\
 Age (Gyr) & -&8.30& 7.24 \\
 $T_{\mathrm{eff}}$ (K) &$4750\pm250\tablenotemark{b}$&  4800 &  4735 \\
 $\log g$ (dex) &$3.5\pm0.5\tablenotemark{b}$&   3.26 &   3.27 \\
 $R/{\mathrm R}_{\odot}$ &$3.97\pm0.06\tablenotemark{a}$  &3.94 &  4.08 \\
 $L/{\mathrm L}_{\odot}$ & - &7.39 &   7.22 \\
 $Z_{i}$ & - &0.015 &  0.022 \\
$X_{i}$ & - & 0.69 & 0.69 \\
$[\mathrm{Fe/H}]$ & $0.23\pm0.12\tablenotemark{b}$ &$-0.04$  & $0.13$\\
$r_{cz}/R$ & - &0.15 & 0.14 \\ 
$\alpha_{MLT}$ &- &1.80&1.80 \\
$\Delta \nu\, (\mu$Hz) & $16.96\pm0.03$&$16.97$& $16.93$
 \enddata
\tablenotetext{a}{Calculated by using asteroseismic relations (Sec.~\ref{sec:osc}).}
\tablenotetext{b}{Determined by spectroscopic observations (Sec.~\ref{Sec:spec}).}
\tablecomments{$M/{\mathrm M}_{\odot}$ is the mass of the star, $T_{\mathrm{eff}}$ is the effective temperature,
 $\log g$ is the surface gravity, $R/{\mathrm R}_{\odot}$ is the surface radius, $L/{\mathrm L}_{\odot}$ is the luminosity, $Z_i$ is  the initial metallicity, $X_i$ is the initial hydrogen abundance, $[\mathrm{Fe/H}]$ is the iron abundance, $r_{cz}$ is the location of the base of the convective region, $\alpha_{MLT}$ is the mixing-length parameter and $\Delta \nu$ is the large separation obtained by linearly fitting the radial-mode frequencies.}
\label{tab:fitted}
 \end{deluxetable}
 
 In Table~\ref{tab:fitted} we give a comprehensive set of stellar properties for the two best fitting models compared to observations of KIC~4448777.

\begin{figure}[!ht]
  \centering
\includegraphics[height=8cm]{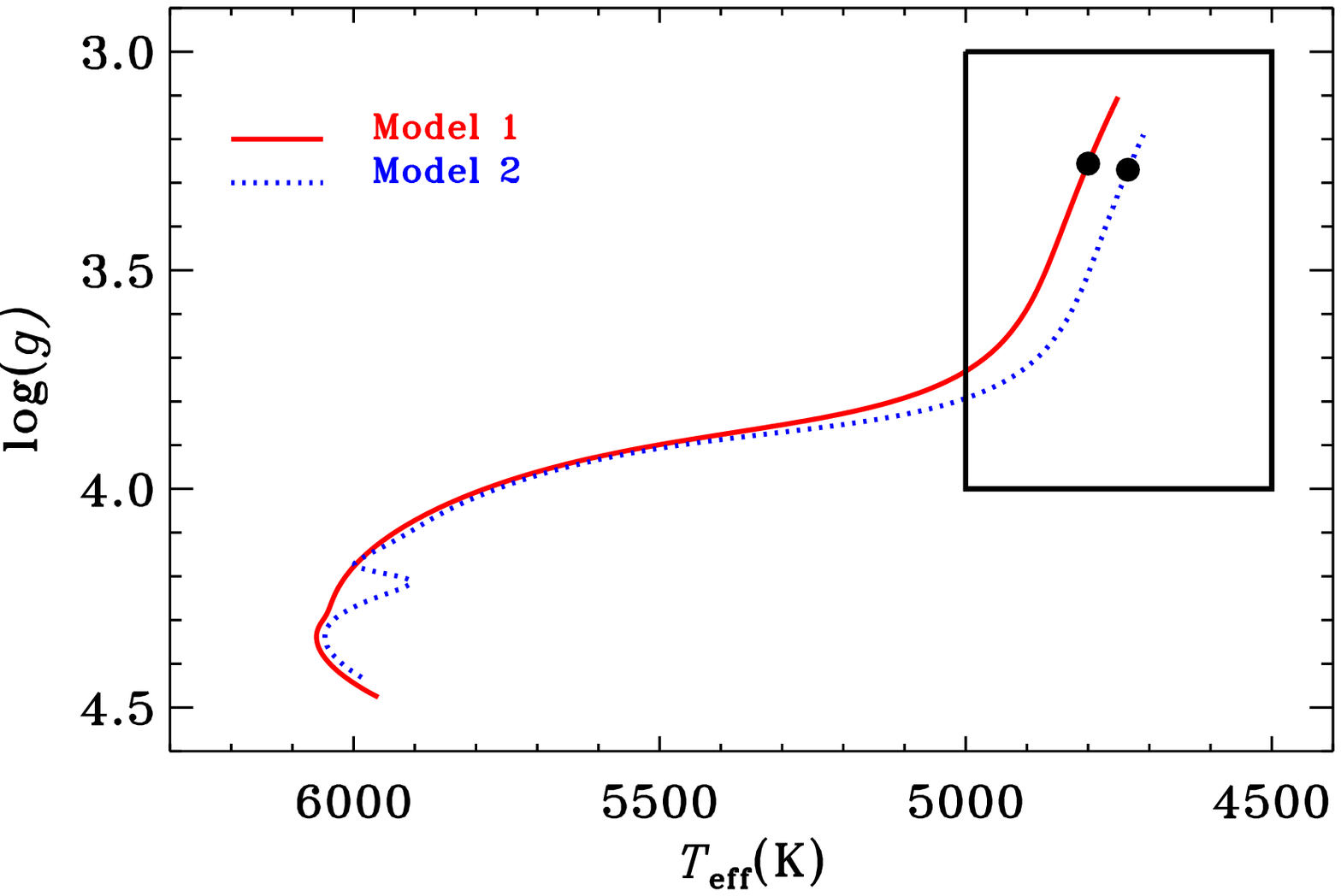}
\caption{Evolutionary tracks plotted in an H-R diagram. Black dots indicate two models which best reproduce the observations. The rectangle indicates the 1~$\sigma$ error box of the
observed
$\log g$ and $T_\mathrm{eff}$.}
  \label{fig:tracks}
\end{figure}

\begin{figure}[!ht]
  \centering
\includegraphics[height=8cm]{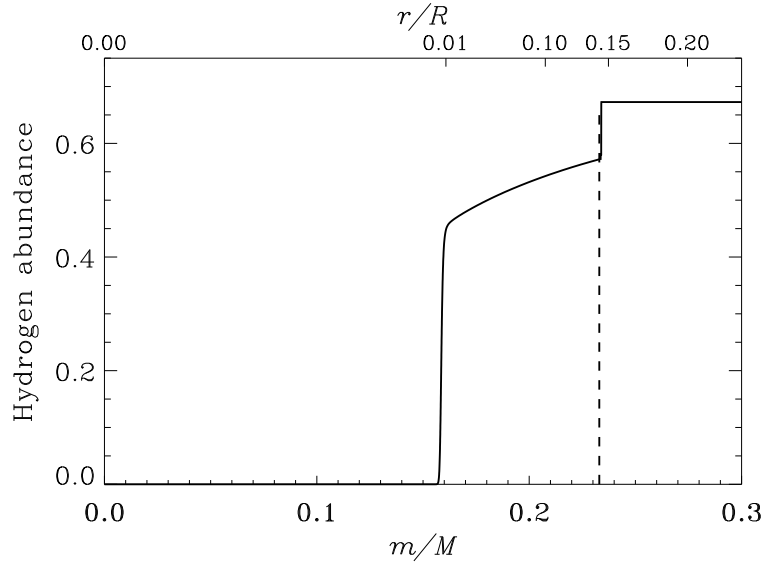}
\caption{Hydrogen content in Model~2 of KIC~4448777. The base of the
convective envelope located at $m_{cz}=0.23M$ and $r_{cz}= 0.14R$ is shown by the dashed line.}
  \label{Xr}
\end{figure}

Fig. \ref{fig:tracks} shows evolutionary tracks plotted in a Hertzsprung-Russell diagram
for the two best-fitting models.
The location of the star in the H-R diagram identifies KIC~4448777 as being at the beginning of the
ascending red-giant branch. It has a small degenerate helium core, having
 exhausted its
central hydrogen and it is in the shell-hydrogen burning phase. The hydrogen abundance as a function of the fractional mass and radius plotted for one of the selected model of KIC~4448777 shows
the extent of the core with a radius $r_c=0.01R$ and the location of the base of the convective zone (Fig. \ref{Xr}). 
The outer convective zone appears to be quite deep, reaching about $r_{cz}\simeq 0.15R$.
 It can be noticed that Model~2, during the main sequence phase, develops a convective core,
which lasts almost to the hydrogen exhaustion at the centre.  The higher metallicity of Model~2, in comparison to Model~1, leads to a high opacity and therefore one would expect a lower luminosity and no convective core in this evolutionary phase. However, in Model~2, the  quite low hydrogen abundance determining
a higher mean molecular weight
acts to increase the luminosity, pushing again to develop a convective core.

\begin{figure}
\centering
  \includegraphics[height=8cm]{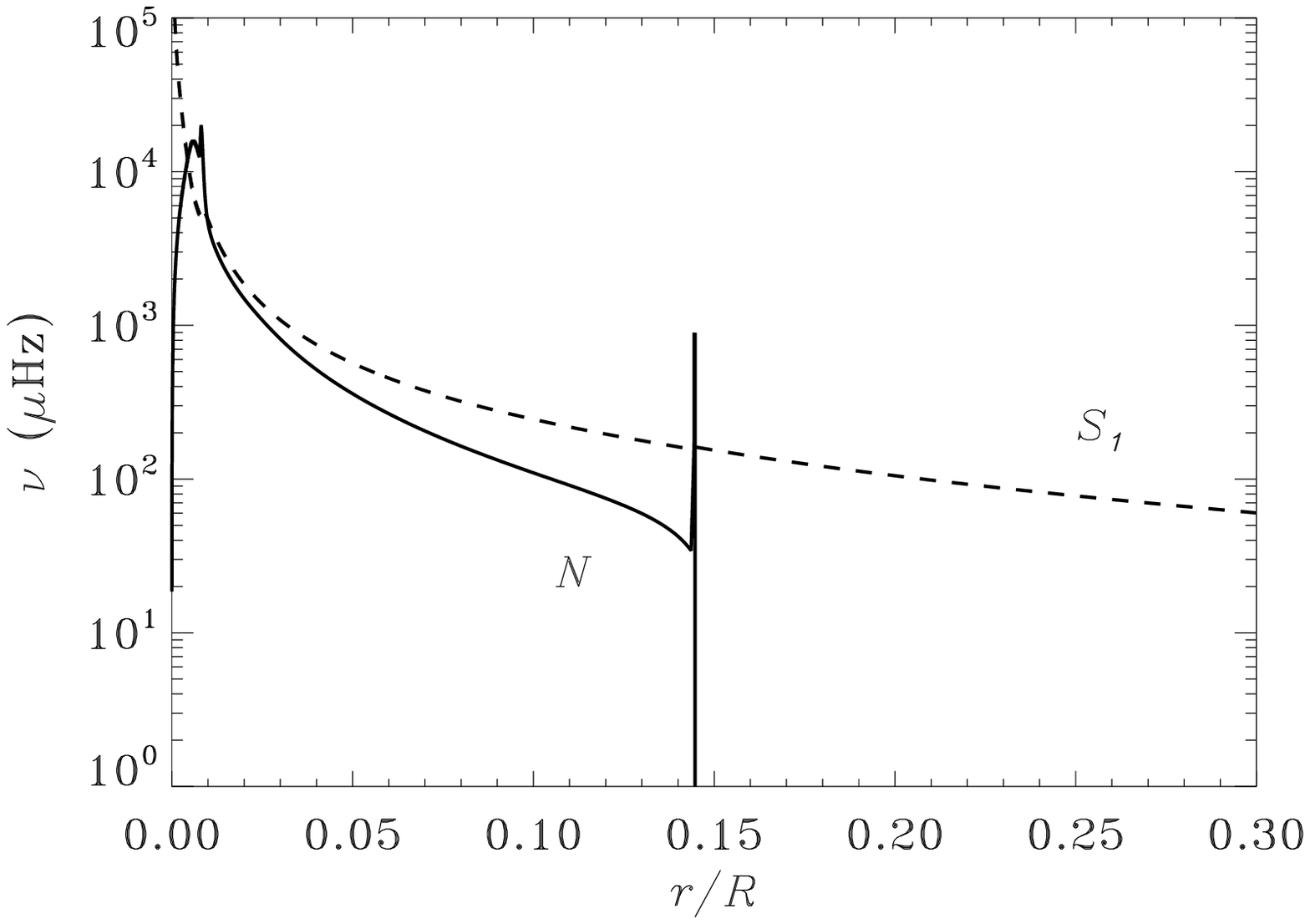}
\caption{Propagation diagram from the center to $ r=0.3\,R$ for Model~1.
  The solid black line represents the buoyancy frequency $N$.
 The dashed line represents the Lamb frequency $S_{l}$ for $l=1$.}
\label{prop} 
\end{figure}

 As shown in the propagation diagram obtained for Model~1 in Fig. \ref{prop}, the huge difference in density between the core region and the convective envelope, causes a large value of the buoyancy frequency in the core, determining well-defined acoustic and gravity-wave cavities, with modest interaction between p and g modes.

\begin{figure}
  \centering
  \includegraphics[height=8cm]{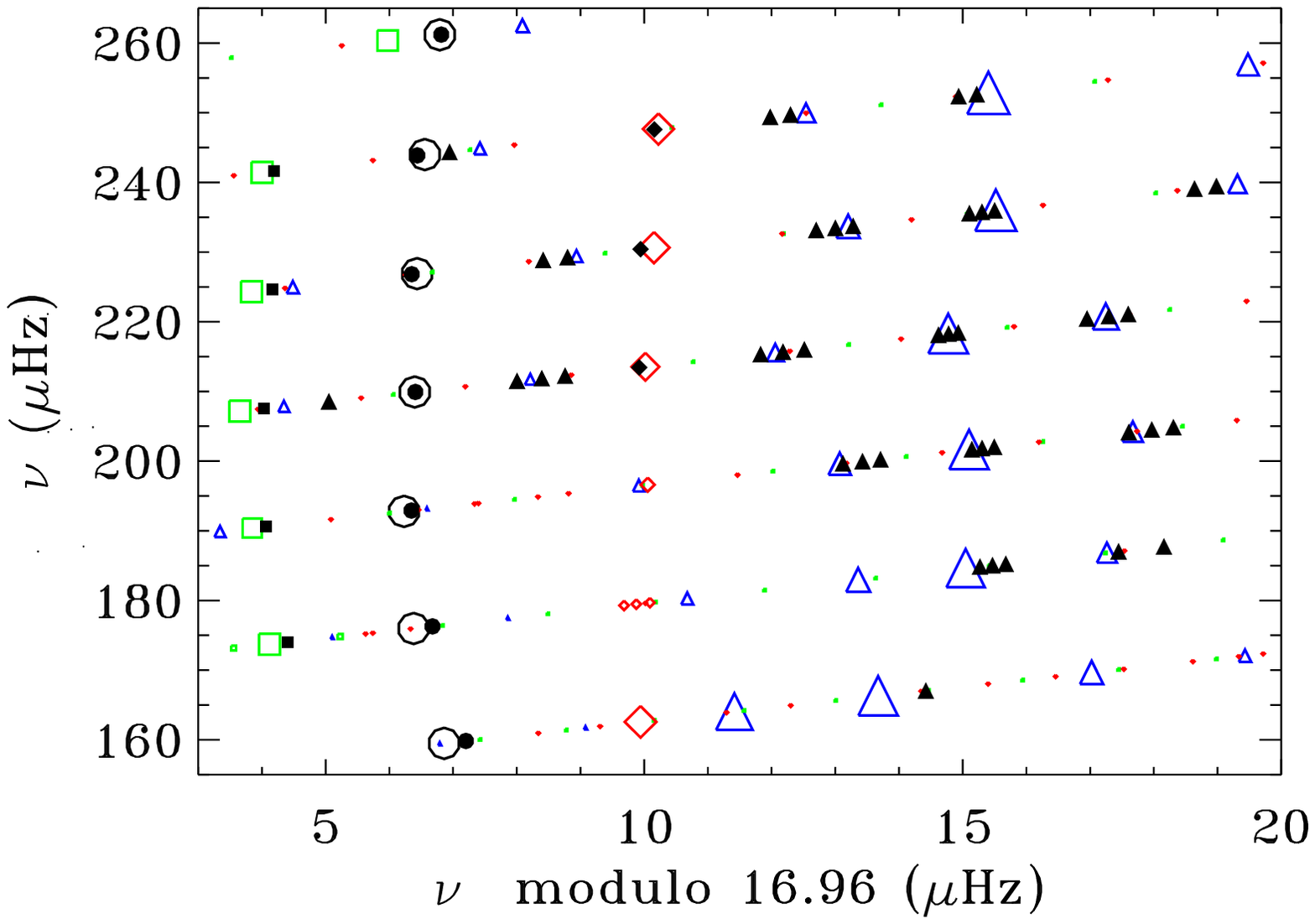}
 \includegraphics[height=8cm]{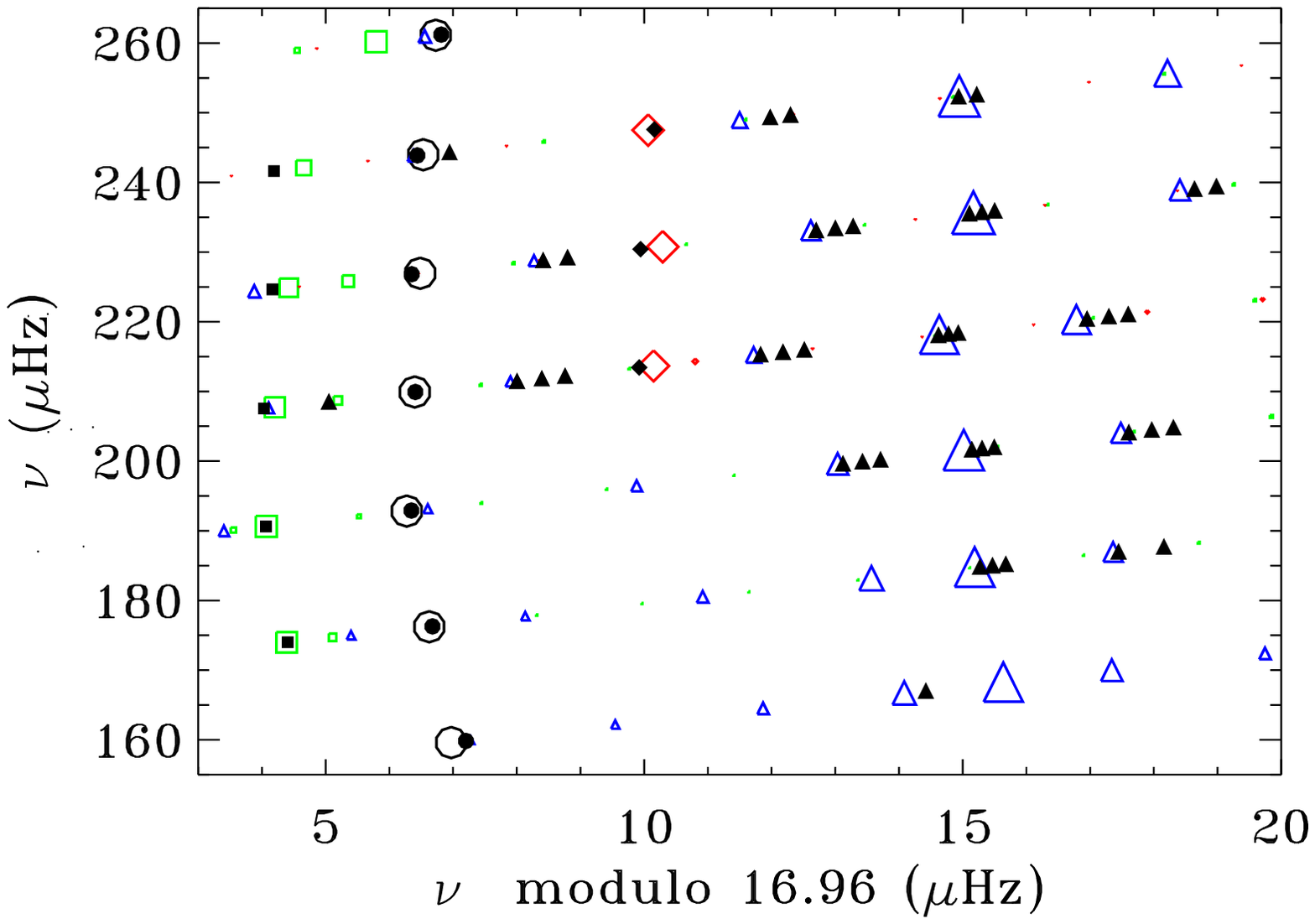}
  \caption{\'Echelle diagrams for Models~1 (upper panel) and 2 (lower panel) of Table \ref{tab:fitted}. The filled symbols show the observed frequencies. The open symbols show computed frequencies. Circles are used for modes with $l=0$, triangles for $l=1$, squares for $l=2$ and diamonds for $l=3$. Observed splitting of $l=1$ modes can be seen as triplets or doublets of black triangles. The size of the open symbols indicates the relative surface amplitude of oscillation of the modes.}
  \label{fig:echelle}
\end{figure}

Figure \ref{fig:echelle} shows the \'echelle diagram obtained for the two models.
The results show,
as explained in previous sections, that the observed modes are
$l=0$ pure acoustic modes, and $l=1,2,3$ g-p mixed modes.  Several non-radial mixed modes have a very low inertia, hence they propagate in the low-density region, namely the acoustic cavity and behave like p modes.
Most of the $l=1$ mixed modes have a quite high inertia, which means that they propagate in the 
gravity-wave cavity in the high-density region, although the mixing  with a p mode enhances their amplitude and hence ensures that they can be observed at the surface. In the \'echelle diagram these gravity-dominated modes evidently departs from the regular solar-like pattern.

We found that there is an agreement between observed and theoretical frequencies of the two selected models, within 4-sigma errors and with $\chi^2=45$ for Model~1 and $\chi^2=61$ for Model~2; however
we notice that 
 % Model~1 best reproduces the asteroseismic estimates of mass and radius (see Sect. \ref{sec:osc}).
 Model~2 best reproduces the spectroscopic observation of the iron abundance.
 
%%%%%%%%%%%%%%%%%%%%%%%%%%%%%%%%%%%%%%%%%%%%%%%%%%%%%%%%%%%%%%%%%%%%%%%%%%
\section{Results of the asteroseismic inversions}
\begin{figure}[!htb]
  \centering
\includegraphics[width=9.6cm]{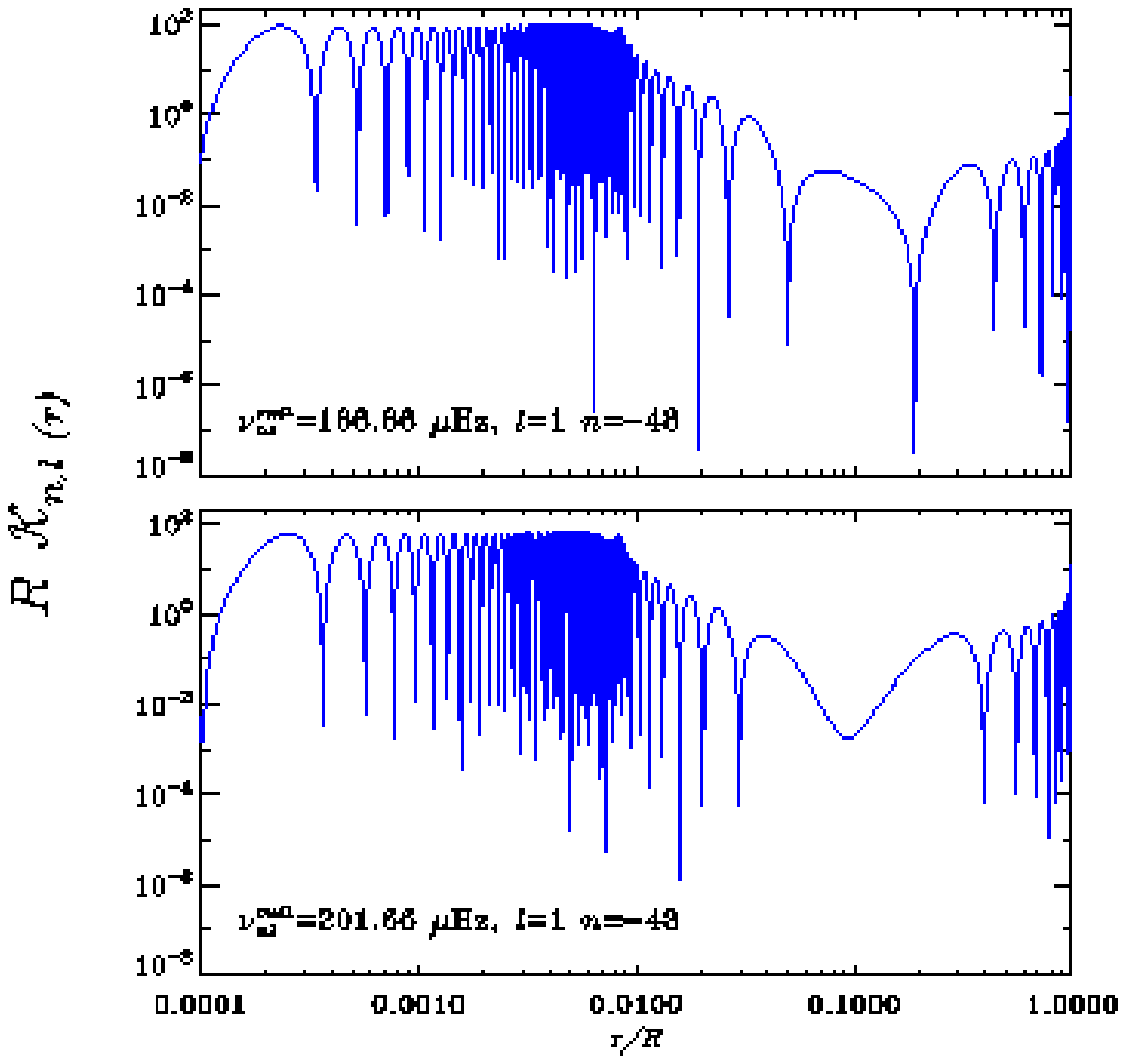}
\hspace{-3cm}
\includegraphics[width=9.6cm]{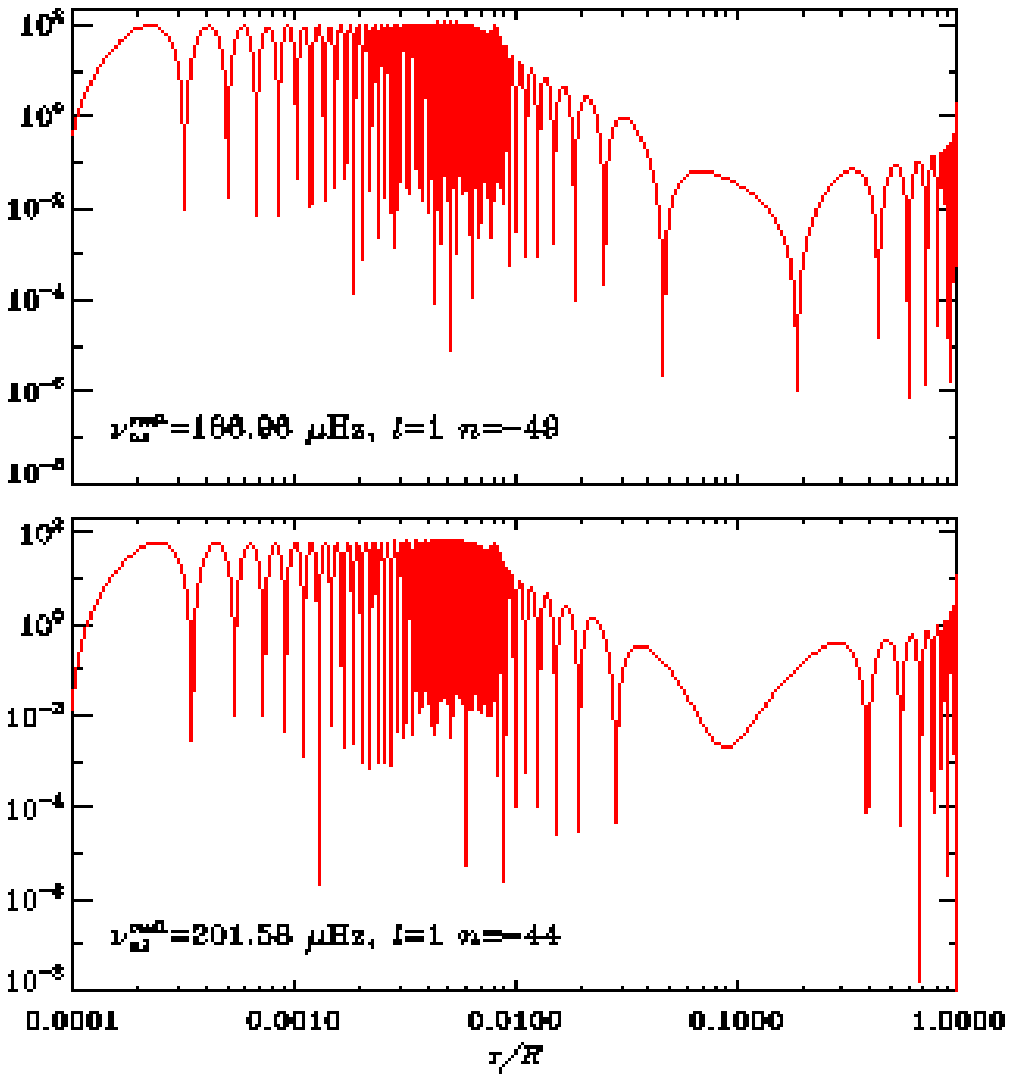}
\caption{Individual kernels calculated for Model~1 (on the left) and Model~2 (on the right) according to
Eq.~\ref{ker} and corresponding to two observed  $l=1$ modes
with frequencies of $187.40\,\mu$Hz and $201.86\,\mu$Hz.
The panels on the top shows a mode with higher inertia than the mode in the bottom panels. 
In each panel the corresponding theoretical oscillation frequency, harmonic degree $l$, and radial order $n$ are indicated.}
  \label{kernl1}
\end{figure}
Once the best model has been selected, it is then possible
to invert  Eq. \ref{eq:rot} following the procedure
 described in the Section~4.
 For this we used the
 14 rotational splittings of the dipole modes given in Table \ref{tab:frequencies}. 

 Kernels calculated for Model~1 and Model~2, 
corresponding to two observed modes with different inertia,
are shown as an example in Fig. \ref{kernl1} \citep[see also][]{goupil96}.
It is interesting to notice that kernels calculated for the two different models, but corresponding to the same frequency, show similar amplitudes in the interior.
\begin{figure}[!htb]
 \begin{center}
\centering
  \includegraphics[height=8cm]{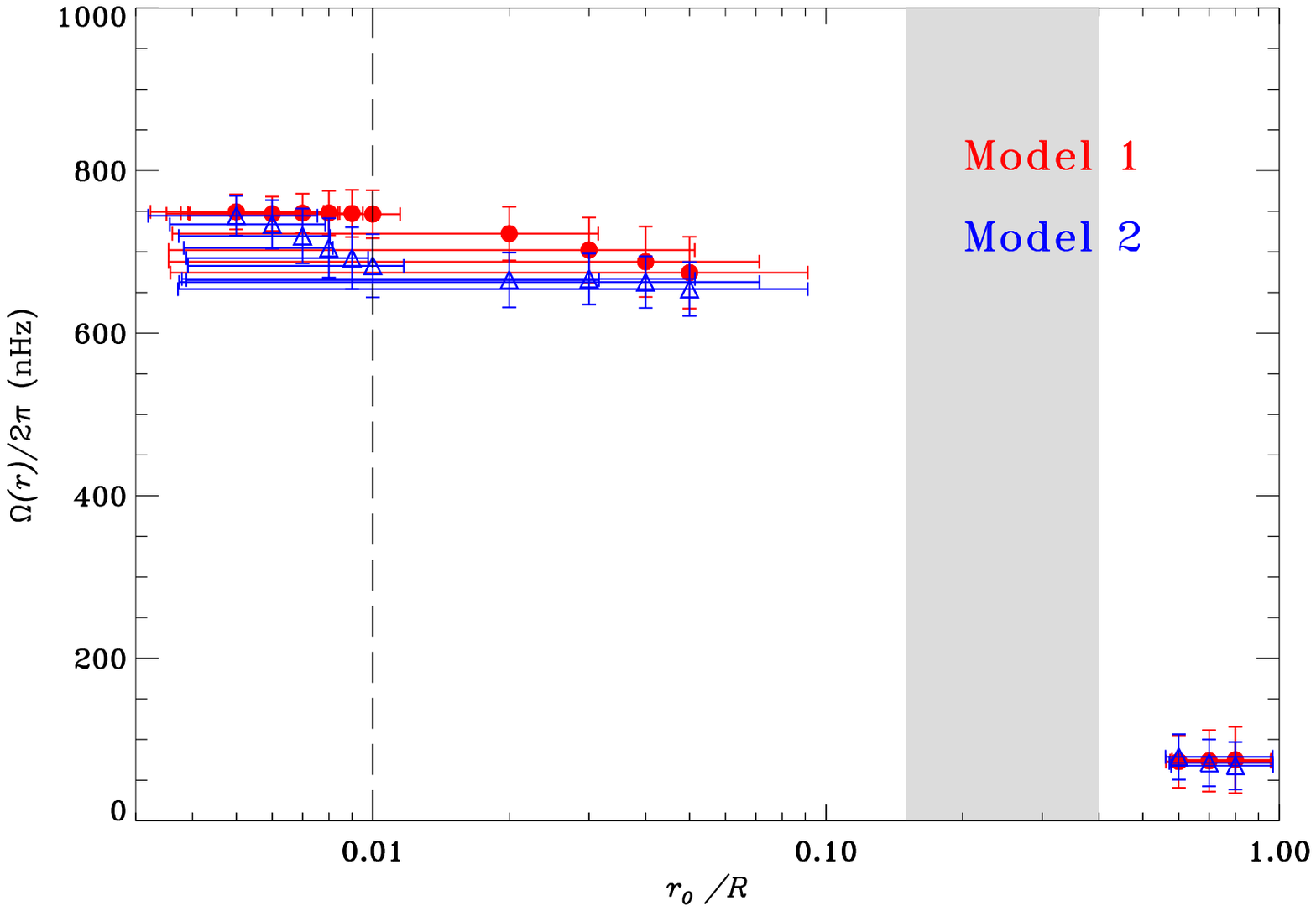}
\end{center}
\caption{Internal rotation of KIC~4448777 at different depths as obtained by the OLA inversion based on the two best-fitting models. Vertical error bars are 2 standard deviations. The dashed line indicates the
location of the inner edge of the H-burning shell.
The shaded area indicates the region inside the star in which it was not possible to determine any solutions.
} 
\label{rot}
\end{figure}

The inferred rotation rate obtained by applying the OLA technique for the two models is shown
in Fig. \ref{rot}, where the points indicate the angular velocity against the selected target radii $\{r_0\}$. %When the averaging kernel present very large sidelobes,  its center of mass results can be shifted relative to the target and the solutions should be considered unreliable.}
The radial spatial resolution is the interquartile range of the averaging kernels and gives a measure of the localization of the solution.   In the probed regions the distance between the center of mass and the target location is smaller than the width of the averaging kernel (see Eq. \ref{cgravity}).

To show more clearly the errors in the inferred internal rotation,
the vertical bars are 2 $\sigma$, $\sigma$ being the standard deviation given by Eq.~10.  Different trade-off
parameters have been used with $\mu=0-10$ to try to better localize the kernels.
%In the case of strong regularization with high trade-off parameter, the averaging kernels show 
%a smooth profile, but the solutions appear less localized.
 A good compromise between localization and error magnification in the solution
 has been obtained using $\mu=0.001$ for both Model~1 and Model~2. The	
   inversion parameter has been	
   chosen by 	
   inverting a known simple	rotational	 
 profile.
% , as can be seen in Figure~\ref{ker2}, which shows the innermost averaging kernels {\b and derived solutions}, obtained
%for different values of the trade-off parameter. 

\begin{figure}
 \begin{center}
\centering
  \includegraphics[height=12cm]{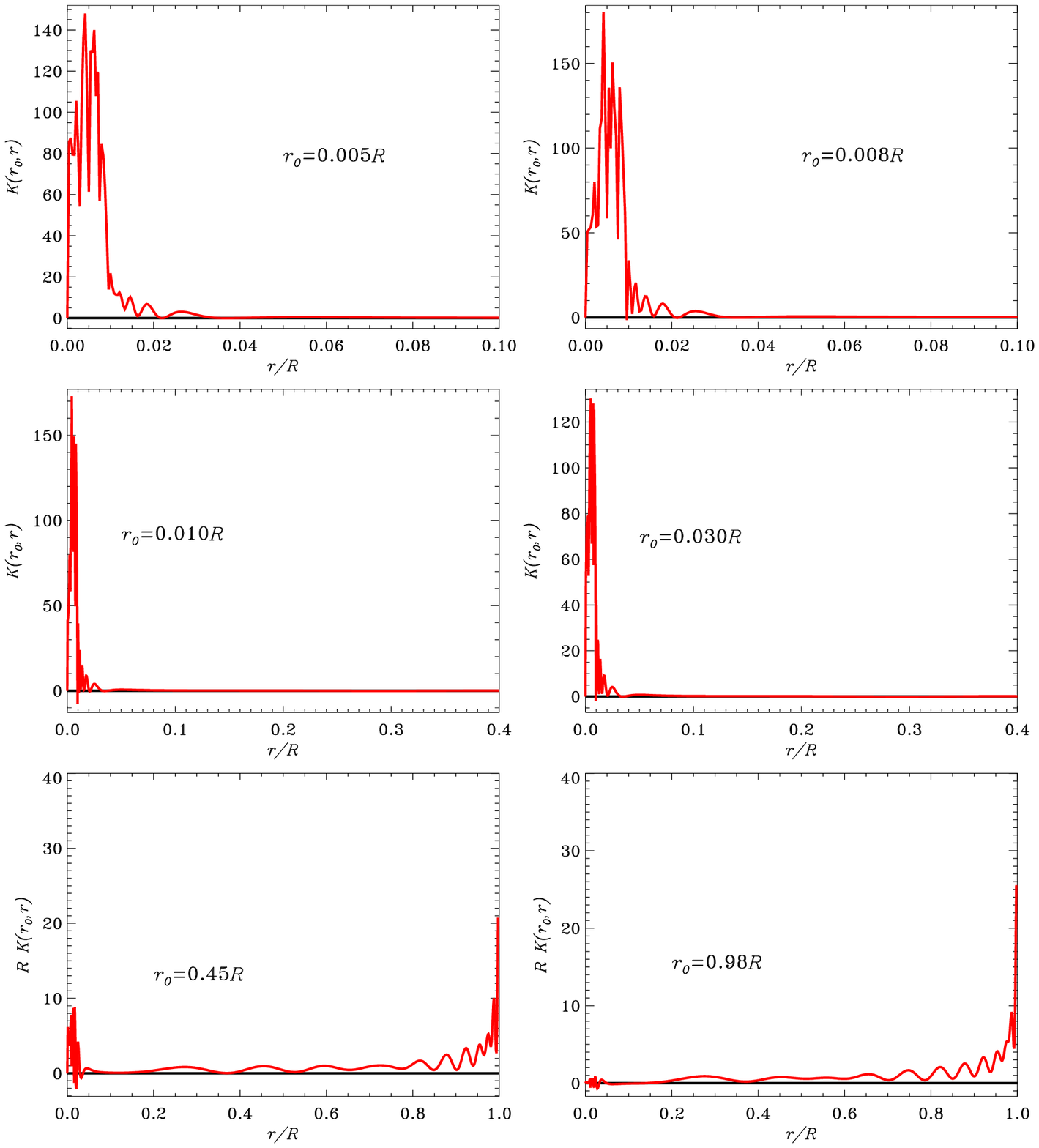}
\end{center}
\caption{OLA averaging kernels localized at several target radii $r_0$ obtained with a trade-off parameter $\mu=0.001$ for the inversion given in Fig. \ref{rot} by using Model~2.}\label{ker1}
\end{figure}

 We were able to estimate the variation of the angular velocity with the radius in the inner interior with a spatial resolution of $\Delta r=0.001R$, thanks to the very localized averaging kernels at different radii.  
 Figure \ref{ker1} shows
OLA averaging kernels localized at several target radii $r_0$ obtained with a trade-off parameter $\mu=0.001$ for the inversion given in Fig. \ref{rot} by using Model~2.

\begin{figure}[!htb]
\centering
  \includegraphics[height=8cm]{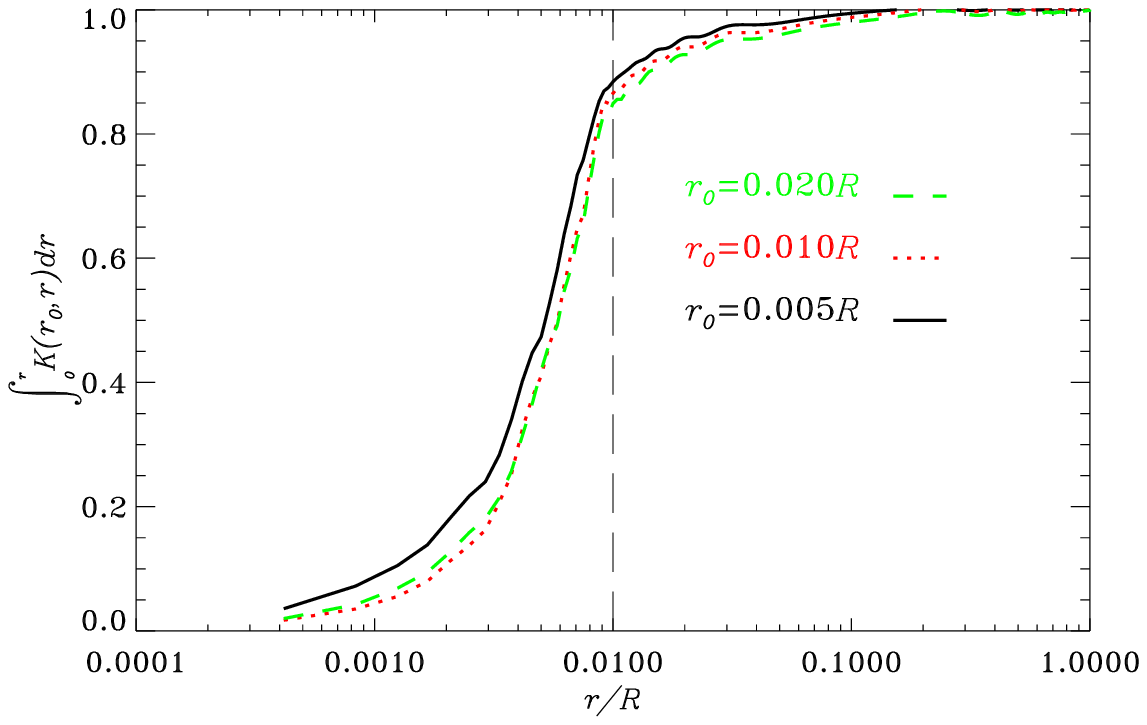} 
 % \vspace{-3cm}
\caption{ Cumulative integrals of the averaging kernels centered at different locations in the inner interior as
  obtained by the OLA inversion using Model~2.  The dashed black line indicates the location of the inner edge of the H-burning shell.}
\label{cum-tuttiola}
\end{figure}

 We find an angular velocity 
in the core at $r=0.005R_{\odot}$ of $\Omega_{c\,\mathrm{OLA}}/2\pi=749\pm11$~nHz with Model 1, well in agreement with the value obtained with Model 2 which is $\Omega_{c\,\mathrm{OLA}}/2\pi=744\pm12$~nHz.  The rotation appears to be constant inside the core and smoothly decreases  from the edge of the helium core through the hydrogen burning shell with increasing radius.
In 
Fig. \ref{cum-tuttiola} we plot the OLA cumulative integrals of the averaging kernels centered at different locations in the inner interior, to show in which region of the star the solutions are most sensitive. 
The cumulative kernels corresponding to solutions centred below and above the H-burning shell look quite similar. The leakage from the core explains the reason why the OLA results show an almost constant rotation in the He core and the H-burning shell.

%, leading to a weighted average value of about $\langle\Omega_{c\,\mathrm{OLA}}/2\pi\rangle=756\pm4$~nHz, combining the results obtained in the core for the two models.
%This shows the limited sensitivity of the inversion results to the core equilibrium stellar model.

We note that it is not possible to find localized solutions for $r_0>0.01R$. Attempts to concentrate solutions above this point return averaging kernels 
which suffer from very large leakage from both the deep core and the superficial layers,  as shown in Fig. \ref{ker1}.

However,
due to the p-mode contributions of certain modes considered, some reliable results can be found above $0.9R$, although with fairly low weight as shown by the kernel centered around $r_0=0.98R$  of Fig. \ref{ker1}.
The angular velocity reaches a mean value in the convective envelope of $\Omega_{s\,\mathrm{OLA}}/2\pi=68\pm22$~nHz with Model 1 in good agreement with $\Omega_{s\,\mathrm{OLA}}/2\pi=60\pm14$~nHz obtained with Model~2.

\begin{figure}
 \begin{center}
\centering
  \includegraphics[height=10cm]{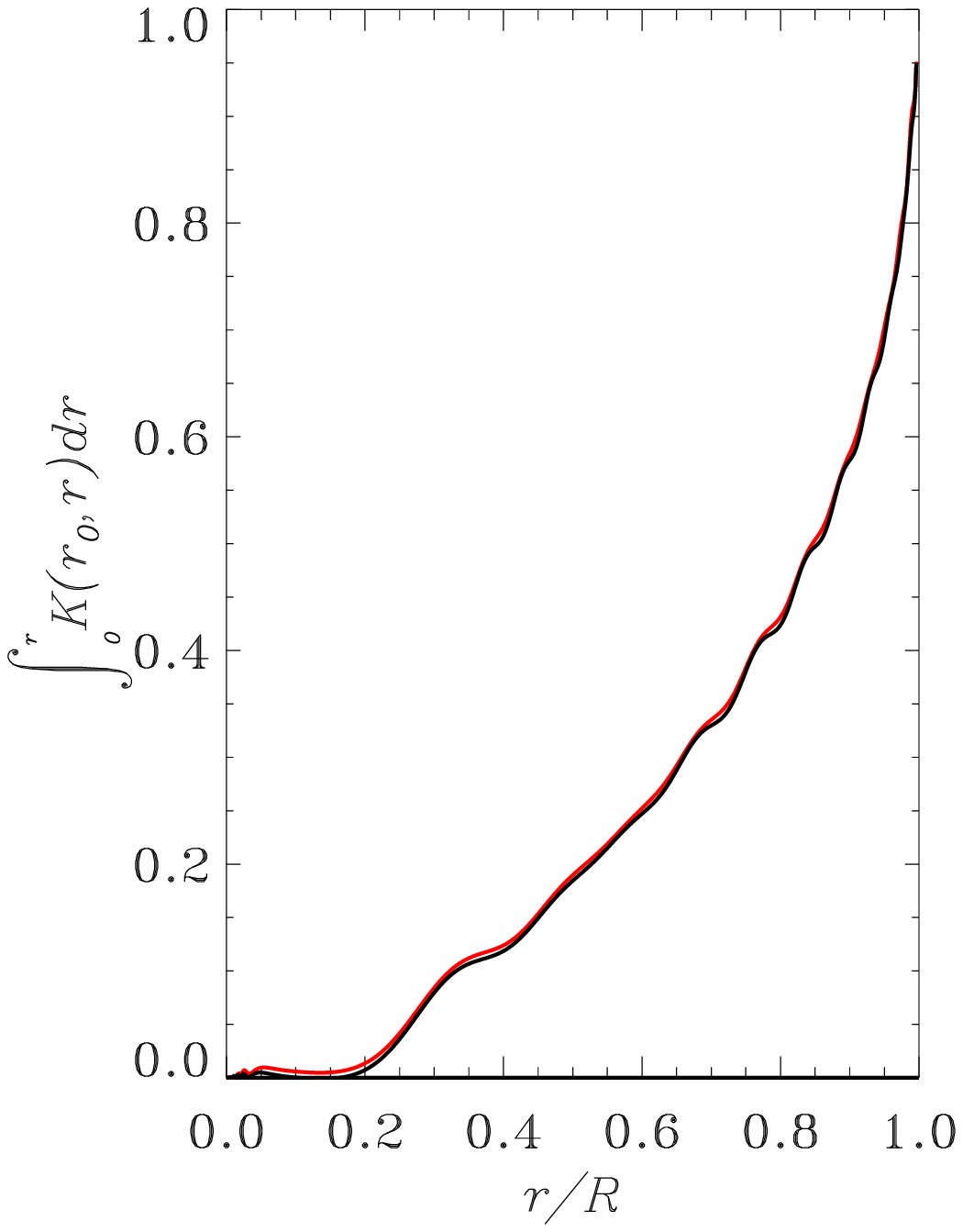}
\end{center}
\caption{Cumulative integral of the surface averaging kernel (centered at $r=0.98R$) 
obtained with the OLA method using Model~1 (red line) and Model~2 (black line).}\label{cum-OLA}
\end{figure}

 The angular velocity value below the surface and the significance of this result 
can be
investigated by considering the cumulative integral of the averaging kernels $\int_0^rK(r_0,r)dr$, in order to understand where the kernels are most sensitive inside the star. 
 Fig.~\ref{cum-OLA} shows that the surface averaging kernels provide a weighted average  of the angular velocity of the layers $r>0.2R$, in most of the convective envelope,
and not an estimate of the rotation at the surface. 
 This is due to the fact that
  the eigenfunctions of the modes considered here are too similar  to one another to build averaging kernels localized at different radii in the acoustic cavity.
 %in our set only few modes
 %are able to probe the acoustic cavity,
 %namely pure p modes or mixed modes with low inertia.

Moreover, Fig.~\ref{cum-OLA} shows that the present set of data does not allow us to appreciate the difference between the cumulative integral of the surface kernels calculated for the two models, hence the results obtained at the surface do not depend on the stellar model chosen.  
The detection of a larger number of modes trapped in the convective envelope would have given the possibility 
%to uniquely
%identify the best model and
 to study 
the upper layers with a higher level of confidence.

%, resulting in a weighted average value of about $\langle\Omega_{s\,\mathrm{OLA}}\rangle=137\pm4$~nHz combining the inversion results obtained at the uppermost layer for the two models.

\begin{figure}
 \begin{center}
\centering
 \includegraphics[height=8cm]{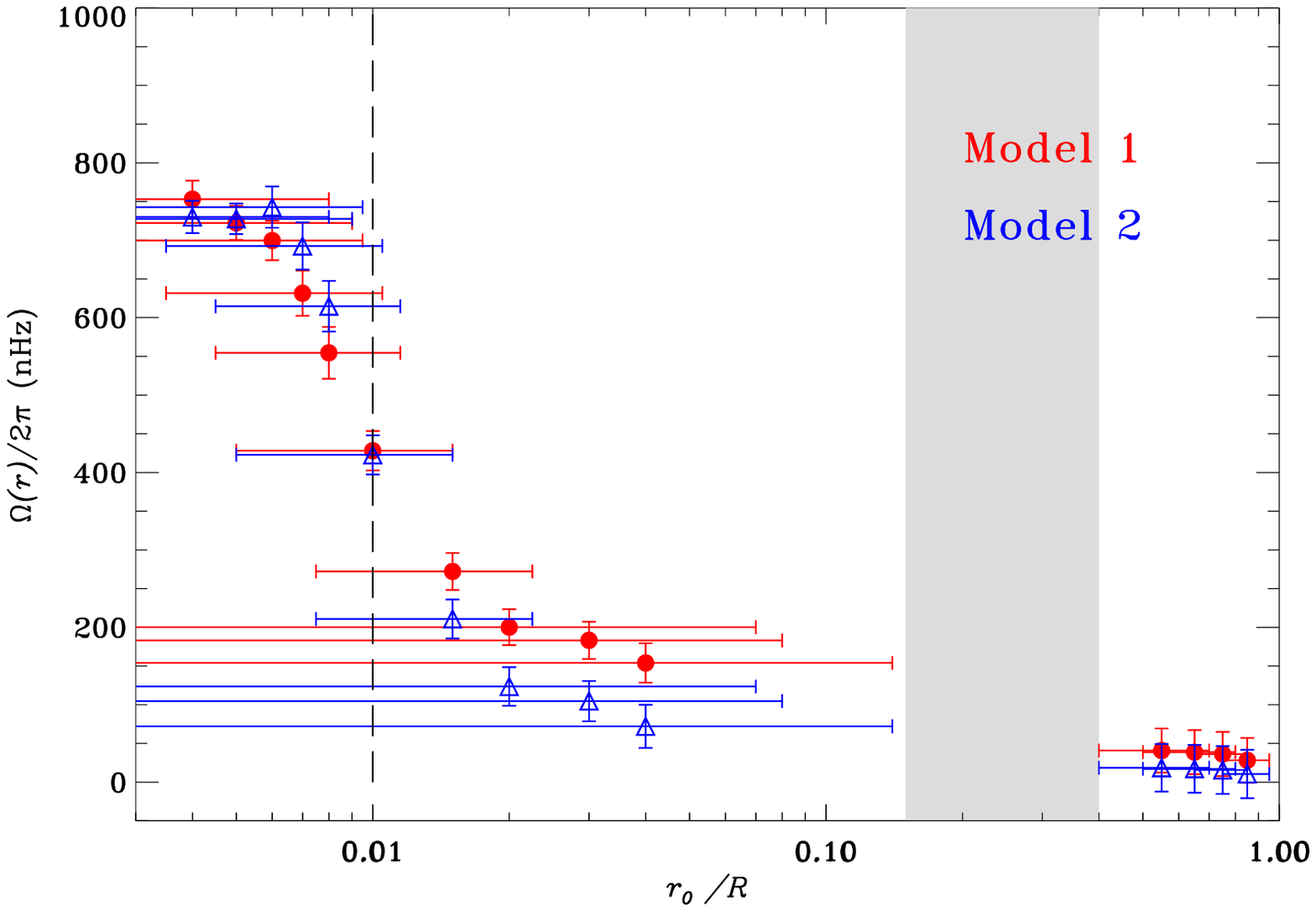}
\end{center}
\caption{Internal rotation of KIC~4448777 at different target radii $r_0$ as obtained by the SOLA inversion for Model 1 and 2. The trade-off parameter used is $\mu=1$.
Vertical error bars are 2 standard deviations.   The dashed line indicates the
location of the inner edge of the H burning shell.
 The shaded area indicates the region inside the star in which it was not possible to determine any solutions.}
\label{rotsola}
\end{figure}

The solutions inferred by the SOLA method, plotted against the target radius, are shown in Fig. \ref{rotsola}. The radial resolution is equal 
to the width of the target Gaussian kernels, while the uncertainty in the solutions is plotted as 2 standard deviations, like for the OLA results.
The values obtained for the angular velocity in the core at $r=0.004R$ are $\Omega_{c\,\mathrm{SOLA}}/2\pi = 754\pm12$~nHz with Model~1 and $\Omega_{c\,\mathrm{SOLA}}/2\pi = 743\pm13$~nHz with Model 2, which are well in agreement with the values obtained by the OLA method.

 %, leading to a weighted average of $\langle\Omega_{c\,\mathrm{SOLA}}/2\pi\rangle=735\pm4$~nHz.
  On the other hand, and differing from the OLA results, above $r=0.01R$ the angular velocity appears to drop down rapidly 
  indicating an almost constant rotation from the edge of the core to the surface.

\begin{figure}
\centering
  \includegraphics[width=12cm]{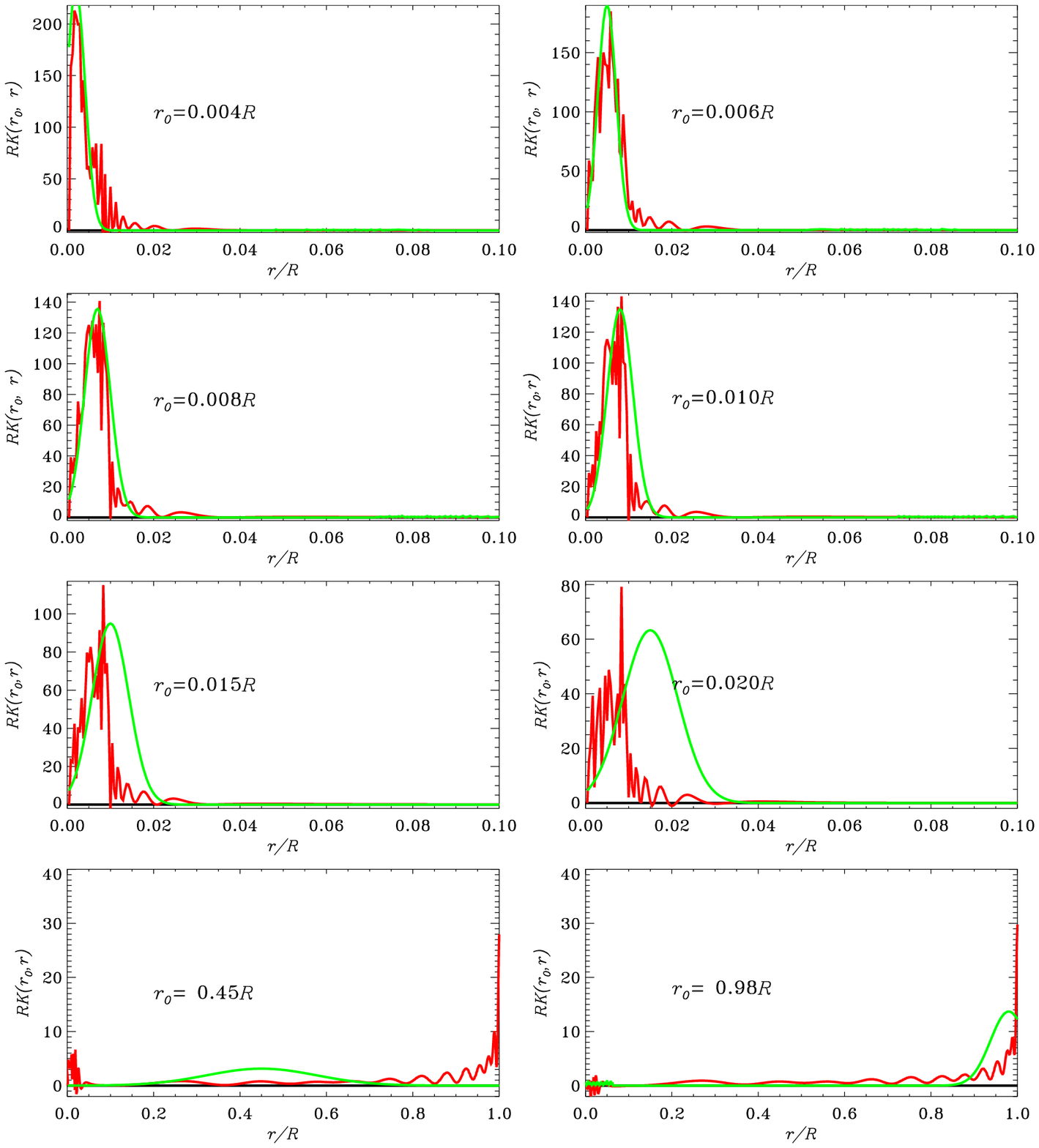}
 
\caption{Averaging kernels (in red) plotted together with the Gaussian target functions (in green) for the SOLA inversion of KIC~4448777, adopting Model~2 and a trade-off parameter $\mu=1$.} 
 \label{kersola}
\end{figure}

The SOLA technique produced reliable results only for
$r<0.01R$, due to the fact that we failed to fit
 the averaging kernels to the Gaussian target function, as required by the method. Figure \ref{kersola} shows averaging kernels and Gaussian target functions for the  solutions plotted in Fig. \ref{rotsola}. Here we used a trade-off parameter $\mu=1$, 
but we notice that the solutions appear to be sensitive to small changes of the trade-off parameter $(0<\mu<10)$ 
 only below the photosphere, with variation up to 20\% in the results. The solutions in the core are not 
 sensitive to the same changes of $\mu$ and the averaging kernels remain well localized.
 It is clear that only solutions related to averaging kernels which are well localized and close to the target Gaussian functions can be considered reliable.

\begin{figure}[!htb]
\centering
  \includegraphics[height=12cm]{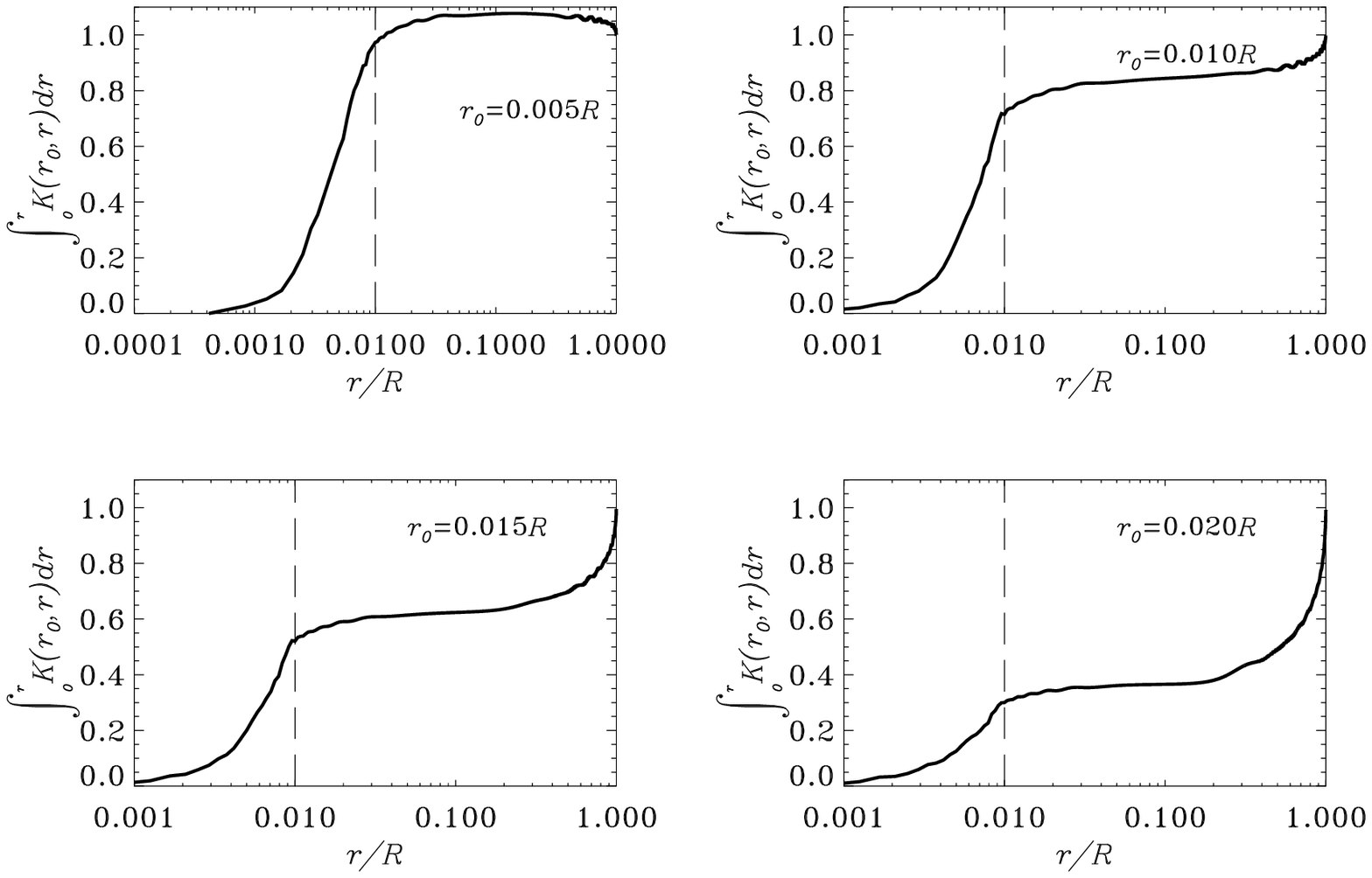} 
 % \vspace{-3cm}
\caption{Cumulative integrals of the averaging kernels centered at different locations in the core 
  obtained with the SOLA method using Model~2.  The dashed line indicates the location of the inner edge of the H-burning shell.}
\label{cum-tutti}
\end{figure}

 In 
Fig. \ref{cum-tutti} we plot the SOLA cumulative integrals of the averaging kernels corresponding to solutions at different locations in the interior. 
We find that the core cumulative kernel is very well localized, and the cumulative kernel for the solution at $r=0.01R$ is localized with a percentage of $70\%$, although contaminated from the layers of the convective envelope. Cumulative kernels for solutions with $r>0.01R$ appear to 
be sensitive to the radiative region for a percentage that is quickly decreasing with increasing target radii, while the contamination from the outer layers appear high. Thus, we can conclude that while the OLA solutions are strongly affected by the core (Fig. \ref{cum-tuttiola}), the SOLA solutions appear more polluted by the signal from the surface layers. Nevertheless, the averaging kernel and cumulative kernel for the SOLA solution at $r=0.01R$ result better localized than the OLA solution indicating that the decrease occurring around the base of the H-burning shell is reliable.

  The angular velocity below the surface at  $r_0=0.85R$ with the SOLA method results to be $\Omega_{s\,\mathrm{SOLA}}/2\pi=28\pm14$~nHz with Model 1 and $\Omega_{s\,\mathrm{SOLA}}/\pi=11
\pm16$~nHz with Model~2. 
  \begin{figure}[!htb]
   \begin{center}
    \includegraphics[height=8cm]{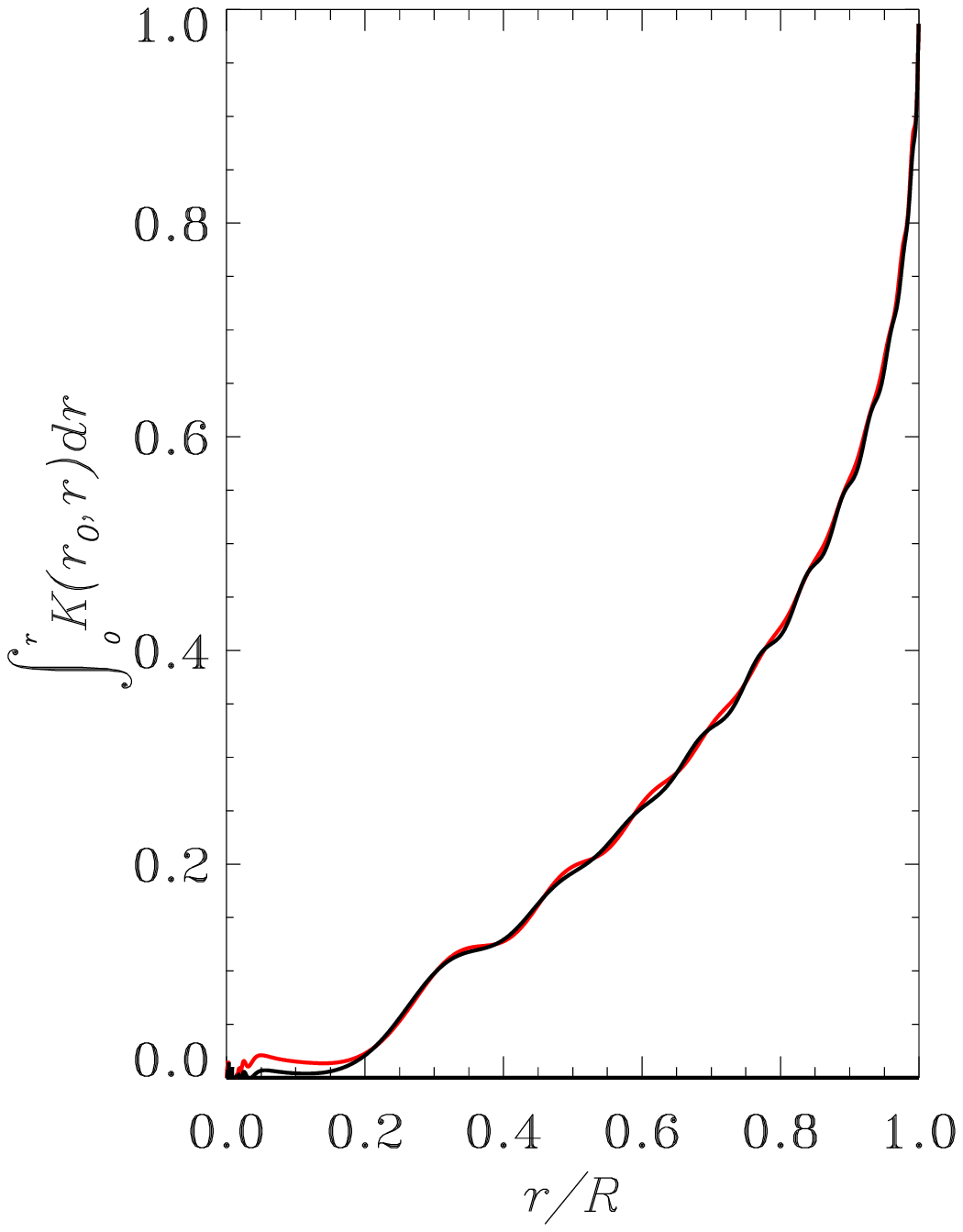}
  \end{center}
  \caption{Cumulative integral of the surface averaging kernel (centered at $r_0=0.85R$) 
  obtained with the SOLA method using Model~1 (red line) and Model~2 (black line).}\label{cum-SOLA}
  \end{figure} 
We can compare the surface cumulative integrals 
  of the averaging kernels $\int_0^rK(r_0,r)dr$
 as obtained for the two inversion methods and  plotted in Figs. \ref{cum-OLA} and \ref{cum-SOLA}.
    We found that the cumulative kernel integral for the 
 near-surface SOLA inversion appears marginally 
 contaminated by the kernels of the regions $r<0.2R$, 
 but the results can still be considered in good agreement with the OLA ones. 
 %This can also be seen by comparing 
 %Figs. \ref{ker1} and \ref{kersola},
 %which show that 
 In the SOLA averaging kernels it was not possible to suppress efficiently the strong signal of the modes concentrated in the core. 
  We conclude that 
  the angular velocity value obtained at the surface by applying the SOLA method represents a weighted average of the angular velocity of the entire interior.  As a consequence, 
  %it is lower than the value obtained with the OLA inversion.
 % For these reasons,
  we think that, in this case, the OLA result
  should be 
  preferred as a probe of the rotation in the convective envelope. 
  
\section{A  strong rotational gradient in the core?} 

 The results obtained in Section 6 raise the question about the possibility of the existence
of a sharp gradient in the rotational profile localized at the edge of the core. 
Evolution of theoretical models  which assumes conservation of angular momentum of the stellar interior predicts that  during the post-main sequence phase, a sharp rotation gradient localized near the H-burning shell,
should form  between a fast-spinning core and a slow-rotating envelope 
\citep[see, e.g.][]{ceillier2013, marques2013}. However, if an instantaneous angular momentum transport mechanism is at work, the whole star should rotate as a solid body. The general understanding is that  the actual stellar rotational picture should be something in between.
 
The occurrence of a sharp rotation gradient in post main sequence stars has been already
investigated by other authors \citep[see, e.g.,][]{deheuvels2014}, with no possibility to get a definitive conclusion.

In order to understand the differences in the inversion results at the edge of the core obtained by using the OLA and the SOLA methods, we tested both techniques by trying to recover 
simple input rotational profiles by computing and inverting artificial rotational splittings.  In order to accomplish this task we used the
forward seismological approach as described in \citet{dimauro2003} for the case of Procyon A.  We computed the expected frequency splittings for several very simple rotational profiles
 by solving  Eq. \ref{eq:rot}, and adopting the  kernels computed from the models used in the present work.  Each set of data includes 14 artificial rotational splittings corresponding to the modes observed for KIC~4448777.  A reasonable error of $7.8$ nHz equal for each rotational splitting has been adopted (see Table \ref{tab:frequencies}). 
The sets of artificial splittings  have been then inverted
following the procedures as described in the above sections. 

 In our tests we used four different input rotational laws:
 a) $\Omega(r)=\Omega_0$; b)  $\Omega(r) = \Omega_c $ for  $r \leq r_c$  $ \Omega(r)=\Omega_e$  for  $r > r_c$; c) $ \Omega(r)=\Omega_0 \cos (2 \pi A \,r)$ where $A$ is a constant;
 d) $ \Omega(r)=\Omega_0+\Omega_1 r+\Omega_2 r^2+\Omega_3 r^3$.
  
\begin{figure}
 \begin{center}
 \includegraphics[width=8cm]{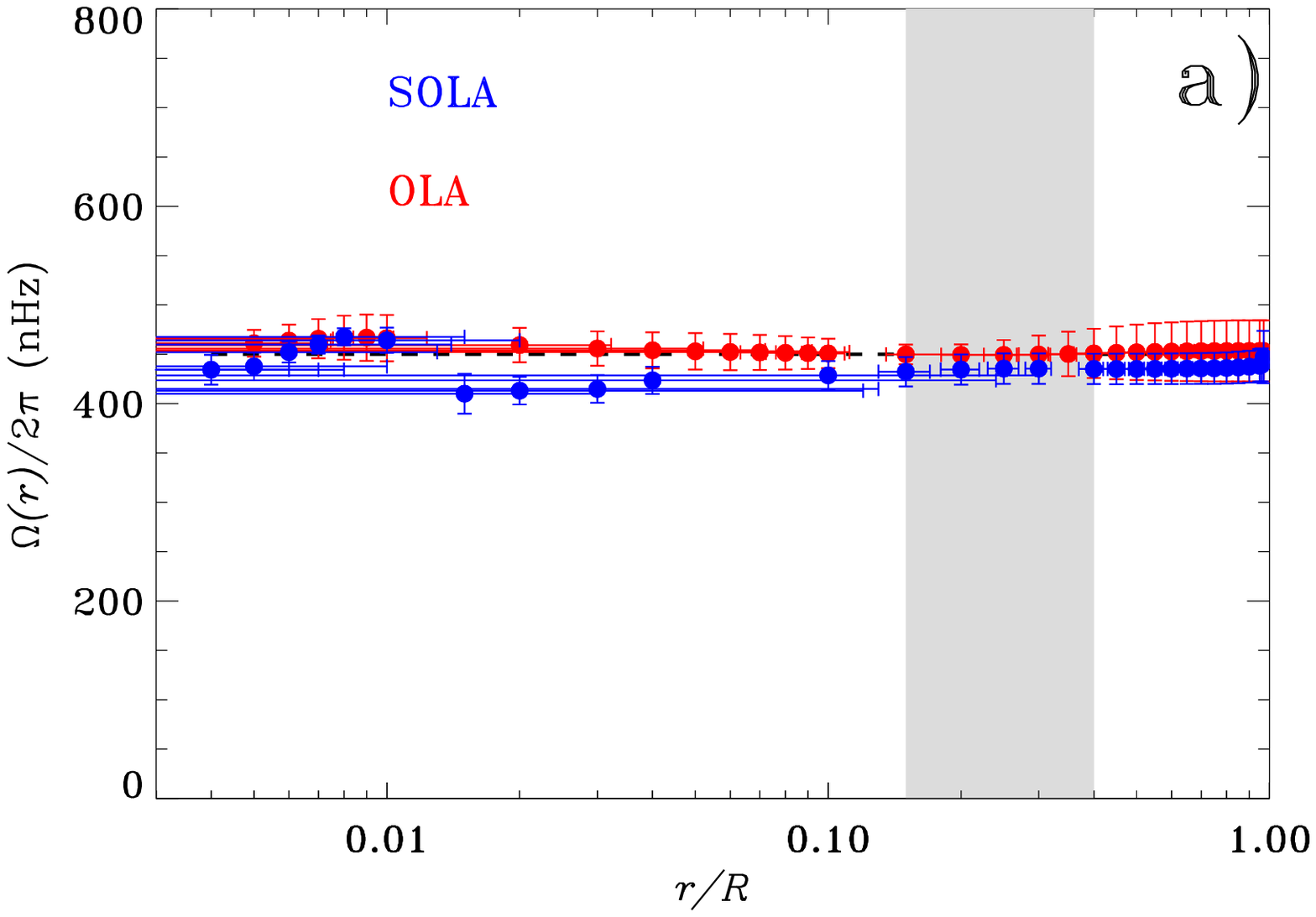}
 \includegraphics[width=8cm]{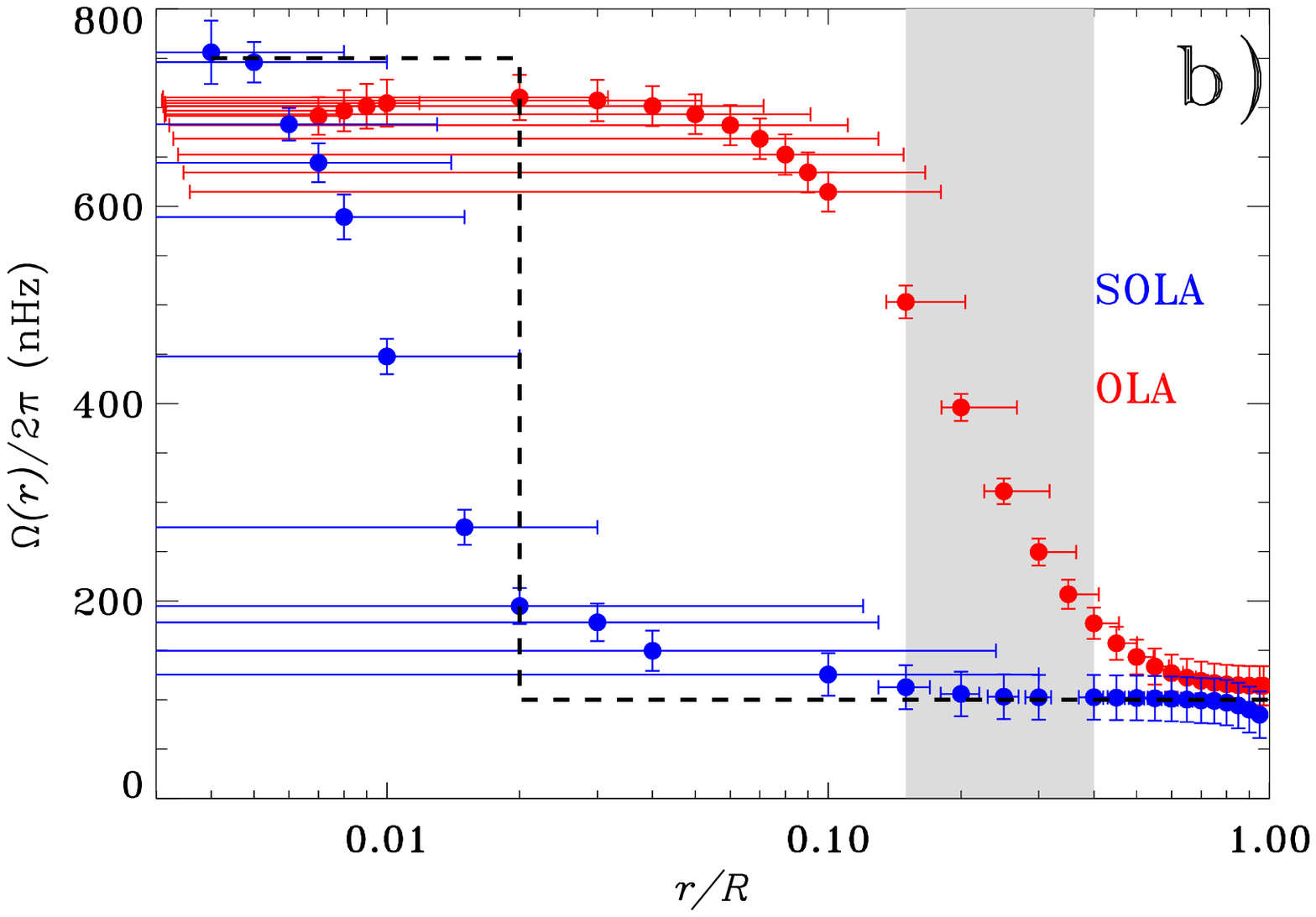}
 \includegraphics[width=8cm]{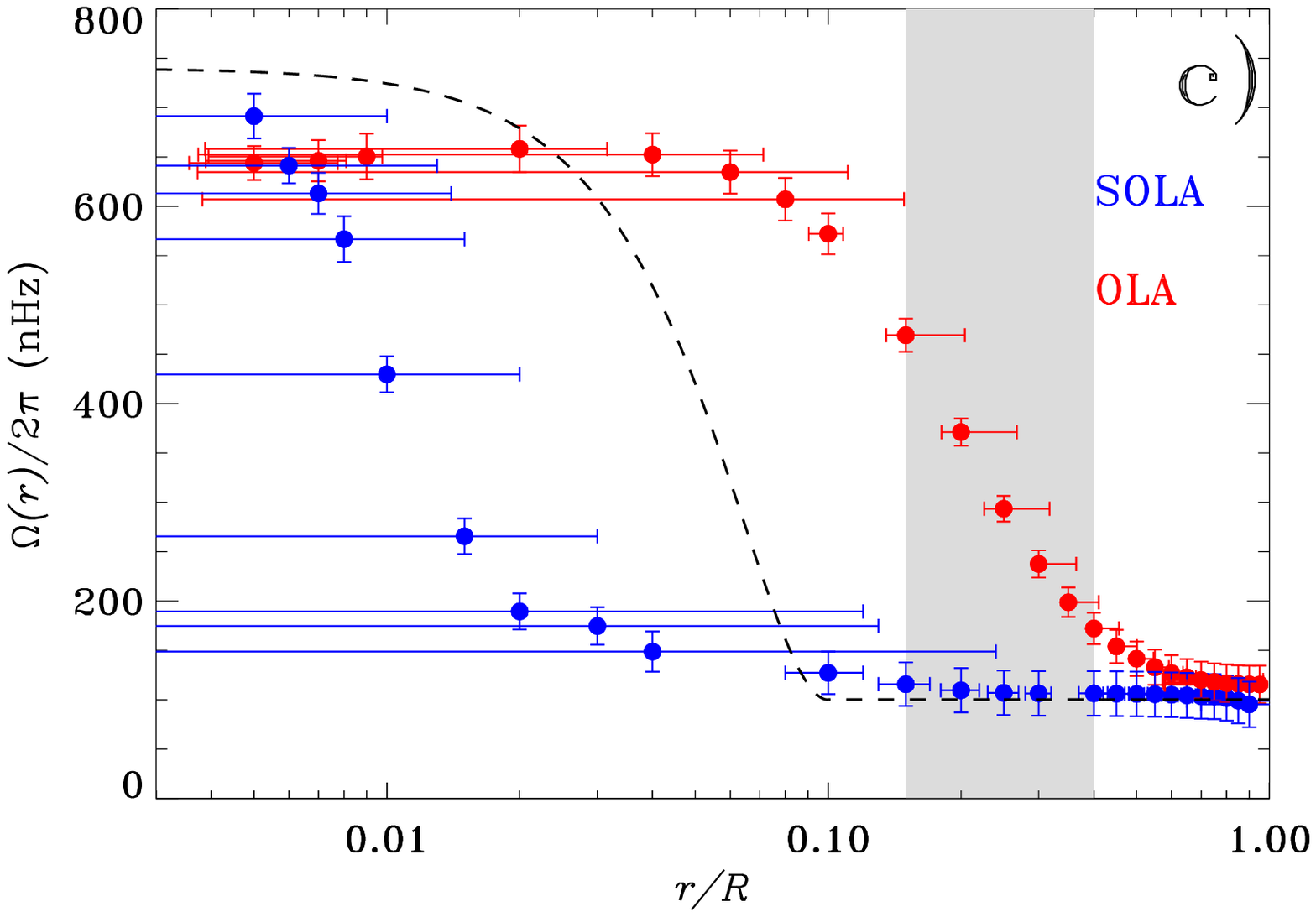}
 \includegraphics[width=8cm]{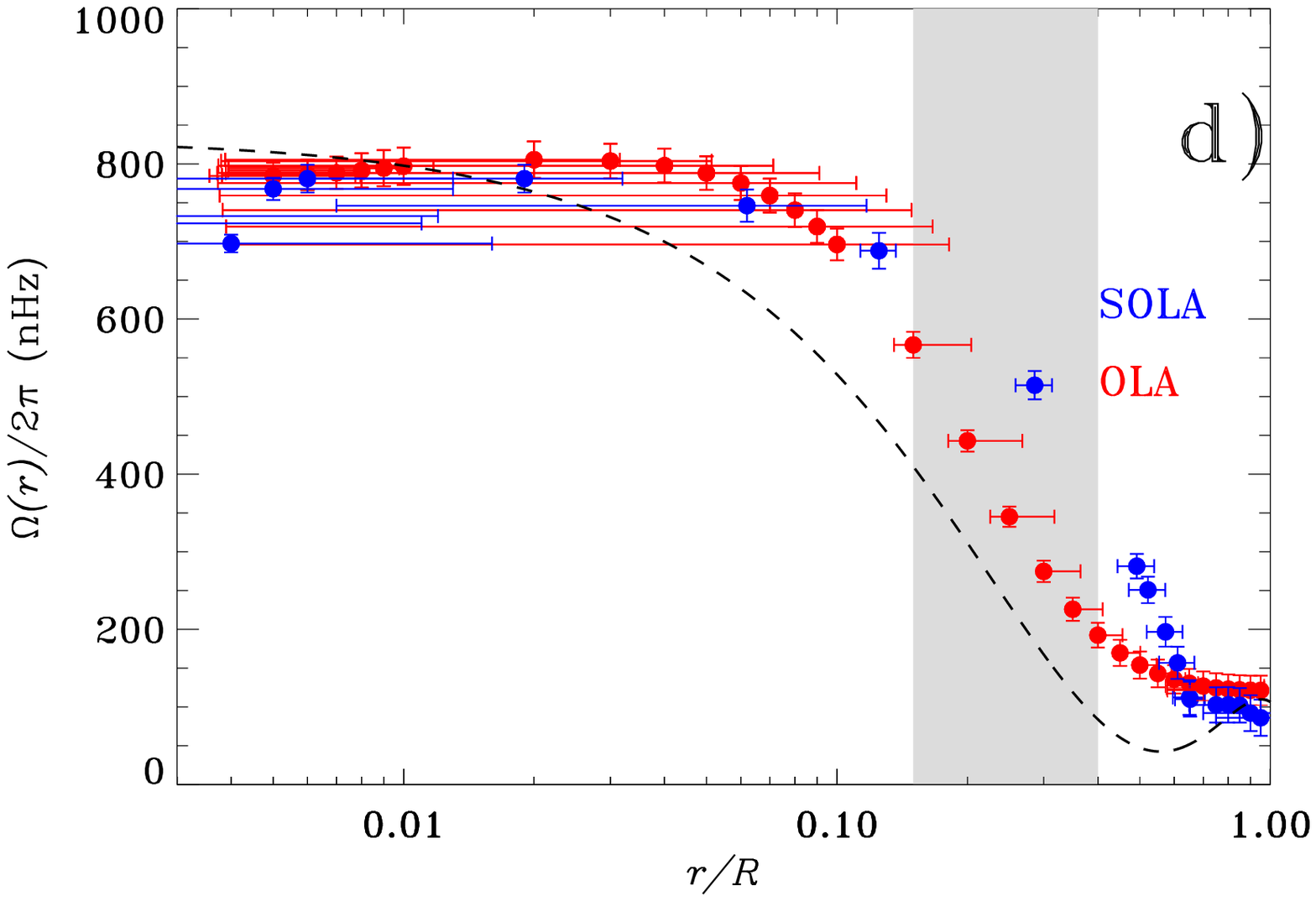}
\end{center}
\caption{Internal rotation of Model 2 at different radii as obtained by the OLA and SOLA inversion of artificial rotational splittings calculated from four simple input rotational profiles  shown by dashed lines. For solution above $r=0.1R$ the horizontal error bars correspond to  
$10\%$ of the radial spatial resolution defined by the averaging kernels. Panel a) shows profile $\Omega(r)/2\pi=450$\,nHz; Panel b) shows profile $\Omega(r)/2\pi=750$\,nHz for $r\leq0.02$ and $\Omega(r)/2\pi=100$\,nHz for $r>0.02$; Panel c) shows profile $ \Omega(r)/2\pi=740 \cos (2 \pi \, 0.5\, r)$\,nHz;  Panel d) shows profile $ \Omega(r)/2\pi=832-3515 r+4992 r^2-2202 r^3$\,nHz. The shaded area indicates the region inside the star in which it was not possible to determine any solutions.}
\label{profile}
\end{figure}

  Figure \ref{profile} shows the input rotational profiles and superimposed the
 results obtained by OLA and SOLA inversions for four of the cases considered with the use of Model~2. Similar solutions have been obtained with Model~1.
  It should be pointed out that, although the panels show inversion results obtained along the entire profile to strengthen the potential of the inversion techniques, as already explained in Sec. 6, the considered set of dipolar modes allows to probe properly only
 the regions where the modes are mostly localized. 

We find that both the OLA and SOLA  techniques are able to well reproduce the angular velocity in the core at $0.005R$ and in the convective zone for $r\geq0.4R$ producing results well in agreement and model independent. The OLA and SOLA techniques well recover rotational profiles having convective envelopes or entire interior which rotate rigidly.  Figures \ref{profile}b), c) and d)  show that both
 techniques are able to measure the maximum gradient strength of an internal rotational profile with a steep discontinuity,  both in the case of decreasing and increasing gradient toward the surface. 
 However the method fails to localize discontinuities with an accuracy better than $0.1R$, due to the progressively increasing errors in spatial resolution at increasing distance from the core.  % if lying between $0.005\leq r\leq 0.1R$, because the modes considered are able to sound only this thin region, as already shown in Sec. 6.
  %However we observe a difference in the results obtained by the two method.
  In the case of the OLA technique the inverted rotational profile decreases smoothly towards the surface even in the case of  a step-like input rotational law.
 On the other hand, with the SOLA inversion we find it more difficult to localize averaging kernels at target radii near the layer of strong gradient of angular velocity.
   See for example results plotted in Figs. \ref{profile}b), c), d), obtained for internal profiles characterized by discontinuities with different slopes.
     
  In addition we notice that the inversion of the observed set of dipolar modes is not able to recover more complicated rotational profiles (as the case Fig. \ref{profile}d).

We can conclude from our tests that with a small set of only dipolar modes, we  have sufficient information to study
  the general properties of the internal rotational profile of a red giant, mainly  the maximum gradient strength  and, with some uncertainty, also the  approximate  radial location
  of the peak gradient.
  With the actual set of modes we are not able to distinguish between a smooth or a sharp rotation gradient inside our star.

\section{Internal rotation by other methods}

As we have pointed out in the above section, asteroseismic inversion of a set of 14 dipole-mode rotational splittings enables an estimate of the
angular velocity only in the core and, to some degree, in some part of the radiative interior and in the convective envelope of the red-giant star.

Here we explore the possibility to compare our inversion results and to get additional conclusions on the rotational velocity of the interior by applying different methods. 
This can be achieved by separating in our splittings data the contribution due to the rotation in the radiative region from that of the convective zone.

Recently, \citet{goupil2013} proposed a procedure to investigate the internal rotation of red giants from the observations. They found that an indication of the
average rotation in the envelope and the radiative interior can be obtained
by estimating the trapping of the observed modes through the parameter
 $\zeta=I_g/I$, the ratio between the inertia in the gravity-mode cavity and the total inertia (see Eq.~\ref{inertia}). 
 
 Because of the sharp decrease of the Brunt-V\"ais\"al\"a frequency at the edge of the H burning  shell, the gravity cavity corresponds to the radiative region, while
the convective envelope  essentially corresponds to the acoustic resonant cavity (see Fig. \ref{prop}).
% This parameter quantifies
%the gravity character of the mode, so that
%for $\zeta=0$ the mode is nearly a p mode mainly concentrated in the acoustic cavity, %while for $\zeta=1$
%the mode is dominated by the gravity character and it is mainly concentrated in the central regions.
 Thus, \citet{goupil2013} demonstrated that, for $l=1$ modes,  Eq.~\ref{eq:rot} can be written:
\begin{equation}
2\pi\delta\nu_{n,l}=0.5
 \zeta\Omega_{g}+(1-\zeta)\Omega_{p}=\zeta(0.5\Omega_{g}-\Omega_{p})+\Omega_{p},
\label{linear}
\end{equation}
where $\Omega_{g}$ is the angular velocity averaged over the layers enclosed within the radius  $r_{g}$ of the gravity cavity,
 while $\Omega_{p}$ is the mean rotation in the  acoustic mode cavity.
Equation \ref{linear} shows that a linear relation approximately exists between the observed rotational splittings and the trapping of the corresponding modes.
  
%\begin{equation}
%\langle\Omega_{g}\rangle=\frac{\int_{0}^{r_{cz}} {\cal K}_{i}(r) \Omega(r)dr}{I_g},
%\end{equation}
%while $\langle\Omega_{env}\rangle$ is
%\begin{equation}
%\langle\Omega_{env}\rangle=\frac{\int_{r_{cz}}^{R} {\cal K}_{i}(r) \Omega(r)dr}{I-I_g}.
%\end{equation} 

\begin{figure}[!ht]
  \centering
    \includegraphics[width=10cm]{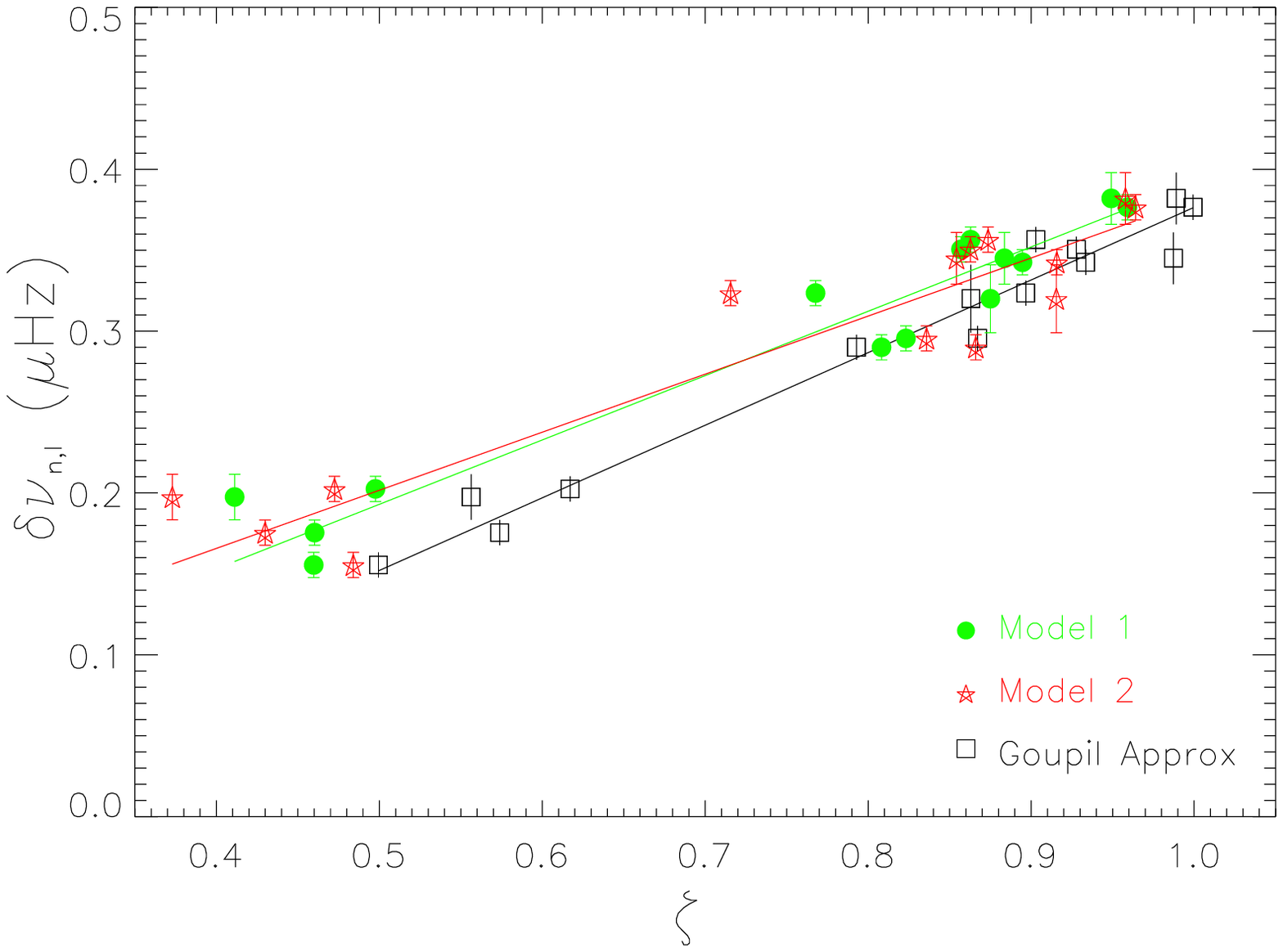}
  \caption{Observed splittings (Model 1: green bullets, Model 2: red stars, Goupil's approximation: black squares) plotted as function of the parameter $\zeta$, which indicates the trapping of the modes. The solid lines shows a linear fit
  of the type $\delta\nu = (a\zeta + b)$.}
   \label{zeta}
  \end{figure}

 The parameter $\zeta$ has been computed for both Model 1 and Model 2 from the relevant eigenfunctions. In order to ascertain the model independence of the results we also computed  $\zeta$  by adopting the approximated  expression given by \citet{goupil2013}, based on the observed values of $\nu_{max}$, $\Delta \nu$ and $\Delta \Pi_1$ (see Sec.~2). Figure \ref{zeta} shows the linear dependence  of the observed rotational splittings on $\zeta$ for both models and for the Goupil's approximation. 
 % and that
%only one splitting lies outside the linear relation. It is not clear if this is due,
%as already noticed by \citet{deheuvels2014} and \citet{goupil2013}, to a wrong estimate %of the splitting or to the rough approximation of the linear dependence. 
%In the following the splitting  $\delta \nu_{n,l} = 0.284~\mu$Hz of the mode %$\nu_{n,l}=252.377~ \mu$Hz was excluded from the treatment of the \cite{goupil2013} method.

%The maximum value of the rotational splittings at $\zeta=1$  gives information
  % on the mean 
  %rotation in the gravity cavity ($0<r<0.15R$):
   %\begin{equation}
   %\langle \Omega_g/2\pi\rangle=2 \delta\nu_{max}=753 \pm 38~\mathrm{nHz}.
   %\end{equation}
   %This value well agrees with that obtained by inversion of splittings in Sec. 6.
   
  We deduced  the mean rotational velocity in the gravity cavity ($0<r<0.15R$), $\Omega_g/2\pi$,  as well as the mean rotational velocity in the acoustic cavity,  $\Omega_p$, 
  by fitting a relation of the type $\delta\nu = (a\zeta + b)$ to the observations, so that $\Omega_g/2\pi = 2(a+b)$ and
  $\Omega_p/2\pi = b$.    We obtained for Model 1 
  %$a= 0.43540819$  and $b = -0.060951108$ so 
  $\Omega_g/2\pi= 762.0\pm31.82$~nHz and $ \Omega_p/2\pi=10.06\pm9.64$~nHz; for Model 2 $\Omega_g/2\pi= 756.48\pm29.86$~nHz and $ \Omega_p/2\pi=38.45\pm9.03$~nHz. The results obtained by adopting Goupil's approximation for $\zeta$: are $\Omega_g/2\pi= 748.90\pm36.30$~nHz and $ \Omega_p/2\pi= - 60.95\pm11.51$~nHz. 
   
It is clear that 
%as it has been noticed by \citet{goupil2013},
   the determination of the mean 
  rotation in the convective envelope of KIC~4448777 is highly difficult by adopting this method, and
  %, because the extrapolation of the value of the rotational splittings at $\zeta=0$ from the linear fit of Fig. \ref{zeta}
  %provides us with a negative mean surface rotation. 
  we can only conclude that the value of $\Omega_p$ is certainly lower compared to the angular velocity of the core.

Another procedure to assess the angular velocity in the interior
can be obtained by searching for the rotation profile that gives the closest match to the observed rotational splittings by performing a least-squares fitting.
To do that, the stellar radius was cut into $K$ regions delimited by the radii
$0=r_0< r_1< r_2<....r_K=R$. Thus, Eq. \ref{eq:rot} can be modified in the following way:
\begin{equation}
2\pi\delta\nu_{i}=\int_{0}^{R} {\cal K}_{i}(r) \Omega(r)dr=
 \sum_{k=1}^{K}S_{i,k}\Omega_k
\label{averaged}
\end{equation}
where $S_{i,k}=\int_k{\cal K}_{i}(r)dr$ and ${\cal K}_i(r)$ are given by Eq. \ref{ker} for each mode $i$ of the set of data used. $\Omega_k$ represents an average value of the angular velocity
in the region $k$.

To assess the angular velocity in the interior we can perform a least-squares fit to the observations by minimizing the $\chi^2$ function:
\begin{equation}
\chi^2=\sum_{i=1}^N\frac{[2 \pi \delta \nu_i-\sum_{k=1}^{K}S_{i,k}\Omega_k]^2}{\epsilon_i^2}
\label{chiav}
\end{equation}

We have explored several cases for small values of $K$ and we found that the result depends on $K$ and on the choice of the values $r_k$ of the boundaries between each region $k$.
Reasonable values  of the reduced $\chi^2$ for the two models  (Model 1:  $\chi^2$ = 5.10, Model 2: $\chi^2$  = 10.6) have been obtained by cutting the %interior of our models in 3 regions, so that $k=1$ corresponds to the region %$r/R\leq0.01$, $k=2$ is for $0.01< r/R\leq0.2$, $k=3$ is for $0.2< r/R\leq1.0$.
interior of our models in 2 regions, so that the region for
 $k=1$ corresponds to radiative region $r/R\leq0.15$, while the region
  for
$k=2$ corresponds to the convective envelope with
 $0.15< r/R\leq1.0$.

%\begin{deluxetable}{lcc}
%\tabletypesize{\scriptsize}
%\tablewidth{0pc}
% \tablecolumns{3}
% \tablecaption{Averaged rotation rates obtained by Least-Squares fit}
 
\begin{figure}[!ht]
   \begin{center}
  \centering
    \includegraphics[height=8cm]{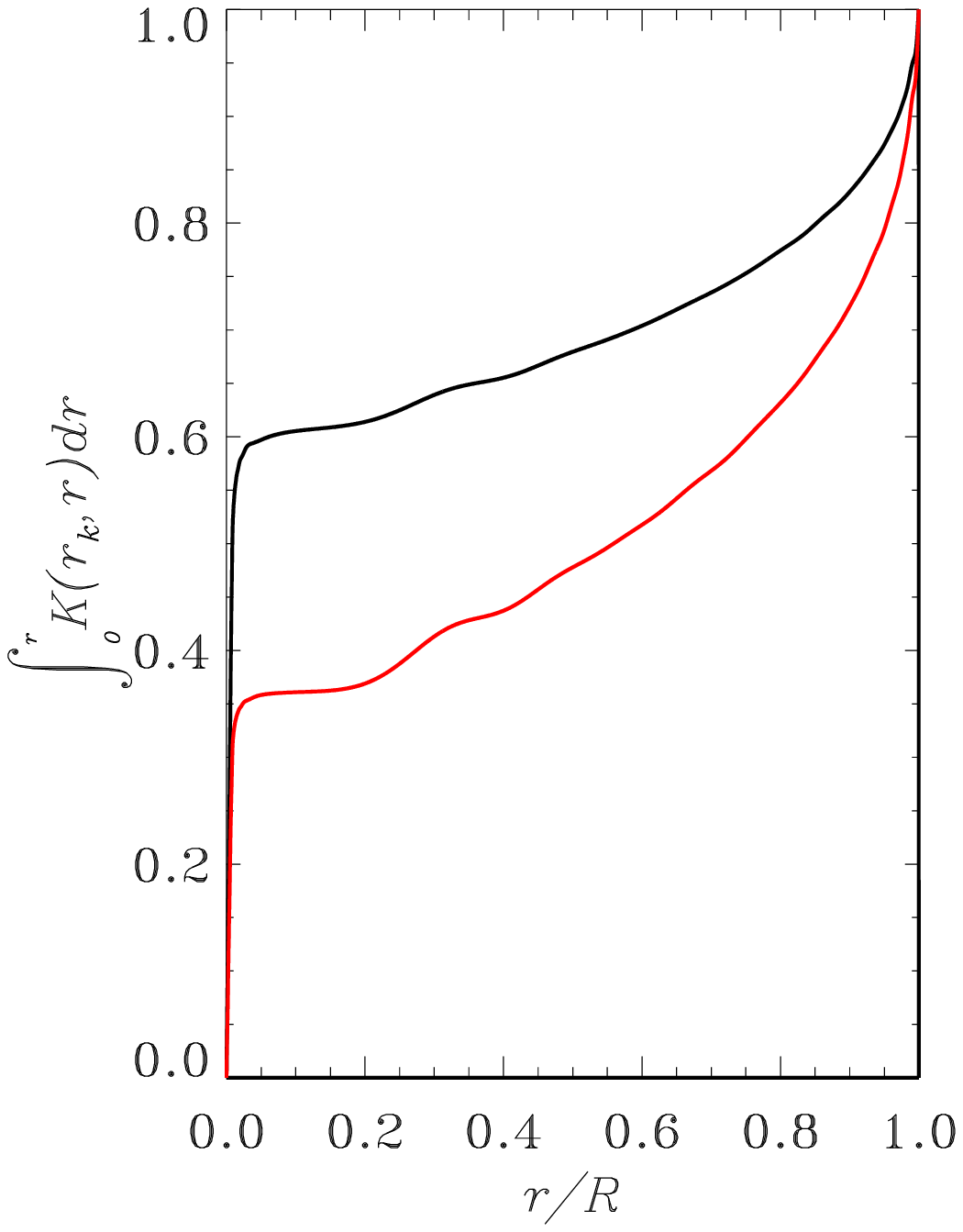}
  \end{center}
  \caption{Cumulative integral of the averaging kernel of the core (black line) and the surface (red line)
  obtained with the LS method using Model~2.}
  \label{cum-LS}
  \end{figure}

% \tablehead{\colhead{} & \colhead{ Model 1}& \colhead{Model 2}}
%\startdata
% $\Omega_1/2\pi$ (nHz) &$744\pm10$ & $735\pm10$\\
% $\Omega_2/2\pi$ (nHz)& $85\pm 5$& $93\pm5$ \\
% \enddata
% \label{omegak}
% \end{deluxetable}
Table \ref{omegak2} lists the values of $\Omega_k$ in the different regions for the two models.
%, shows that the radiative interior of this red giant appears to rotate 8-9 times faster than the convective envelope. 
  Once the least-squares problem has been solved, we can use the coefficients to calculate 
the averaging kernels:
\begin{equation}
K(r_k,r)=\sum_{i=1}^N c_i(r_k){\cal K}_i(r).
\end{equation}
Figure \ref{cum-LS} shows that while the cumulative integrated kernel of the region $k=1$ is quite sensitive to the core, although a contribution from the surface is still present, the cumulative integrated kernel of the region $k=2$ is strongly contemned  by the modes trapped in the radiative interior. 

%In order to understand the role of the 'outlier splitting' shown in Fig.  \ref{zeta}, we repeated the least-squares calculation
%by excluding it from the analyzed set and we obtained the results given in Table \ref{omegak2}. The results in Table \ref{omegak2} show that the outlier has a strong weight for the result at the surface and that precise identification of modes with low inertia trapped in the acoustic cavity is highly important for probing the convective envelope. 
% This explains the reason why $\Omega_p$ is different by $\Omega_2$, since
% we have excluded the splitting at $\nu=284$ nHz in the linear fit.
%It is clear that the least-square fit result does not provide additional information to %what obtained with the OLA and SOLA inversion method.

\begin{deluxetable}{lcc}
 \tablecolumns{3}
 \tablecaption{Averaged rotation rates obtained by Least Square Fit}
\tabletypesize{\scriptsize}
\tablewidth{0pc}
 % excluding $\delta \nu_{n,l} = 0.284~ \mu$Hz of the mode $\nu=252.377~\mu$Hz from the set of data.}
 
 \tablehead{\colhead{} & \colhead{ Model 1}& \colhead{Model 2}}
\startdata 
$ \Omega_1/2\pi$ (nHz) &$765\pm23$ & $746\pm22$\\
  $\Omega_2/2\pi$ (nHz)& $25\pm 8$& $52\pm8$ \\
 \enddata
 \label{omegak2}
 \end{deluxetable}

\section{Conclusion}

In the present paper, we have analyzed the case of KIC~4448777, a star at the beginning of the red-giant phase, for which a set of only 14 rotational splittings of dipolar modes have been identified.
Its 
 internal rotation has been probed successfully by
means of asteroseismic inversion.

We confirm previous findings obtained in other red giants \citep{beck2012,deheuvels2012, deheuvels2014} that the inversion of rotational splittings can be employed to probe the angular velocity not only in the core, but also in the convective envelope. 
 We find that the
helium core of KIC~4448777 rotates faster than the surface at an angular velocity of about 
$\langle\Omega_c/2\pi\rangle=748\pm18$~nHz, obtained as an average value from the SOLA and OLA inversion
techniques. 
 Moreover, we found that the result in the core, well probed by mixed-modes, does not depend on the stellar structure model employed in the inversion.
The value estimated in the core agrees well with those obtained by applying other methods, as the one 
 based on the relation between the observed rotational splittings and the inertia of the modes \citep{goupil2013}, or  the least-squares fit to the observed rotational splittings.

 Our results have shown that the mean rotation in the convective envelope of KIC~4448777
 is $\langle\Omega_s/2\pi\rangle=68\pm22$~nHz,
 obtained as an average of the OLA inversion results for the 
 two 'best-fitting' selected models, indicating 
    that the core should rotate at about $11$ times faster than the convective envelope. 
    For the SOLA inversion solutions, it was not possible to  suppress efficiently the strong signal of the modes concentrated in the core.
    The value of the rotation deduced in the convective zone is compatible with the upper limit measured at the photosphere $\Omega_{ph}/2\pi<538\,\mathrm{nHz}$
    derived from the spectroscopic value of $v\sin i$ (see Sect. 2) and the stellar radius provided by the models. 
 Unfortunately, with few modes able to probe efficiently the acoustic cavity, 
   we have been able to deduce only a weighted average of the rotation in the whole convection zone, but the inversion results appear to not depend on the equilibrium structure model employed.
 Other methods used to determine rotation in the convection zone have provided us with results reasonable in agreement with those obtained by inversion.
 
 Furthermore, we demonstrate that the inversion of rotational splittings can be employed to probe
 the variation with radius of  the angular velocity in the core, because the observed mixed modes
 enable us to build well localized averaging kernels. 
The application of both SOLA and OLA inversion techniques allowed us to
show that the entire core is rotating with a constant angular velocity.  In addition, the SOLA method
 found evidence for an angular velocity decrease occurring in the region $0.007R\leq r \leq0.1R$, between the helium core and part of the hydrogen burning shell,
 %at the inner edge of the hydrogen burning shell, 
 which cannot be better localized,
  due to the intrinsic limits of the applied technique and to the lower resolving power of the employed modes 
in the regions above the core. 
 Thus, although we are not able to distinguish between a smooth or a sharp gradient,
we can determine with good approximation the maximum gradient strength and  the radial position of the peak gradient.

% Our  results provide evidence of a sharp gradient  in the inner rotational profile, localized in the region between $0.005\leq r \leq0.1$, somewhere in between the helium core and the hydrogen burning shell. 
%This result open interesting question on the efficiency of the angular momentum transport mechanism in the evolution from main sequence to red giant phase. 

With the available data, including just a modest number of dipolar modes, it is clearly impossible to infer the complete internal rotation law of KIC~4448777.
In order to resolve the regions above the core, it is necessary to invert
 a set of data which includes more rotational splittings, and in particular of modes with $l>1$ and with significant amplitudes in the acoustic cavity. Such data may become available for other targets or from analysis of longer time-series observations than those considered here.
% In a future work we plan to analyzed the complete set of the four year Kepler observations of the star.

%Once deduced the internal rotational profile of a red giant, we will be able to derive %its angular momentum.
It seems fair to say that, at this stage, the asteroseismic inversions are 
giving  very useful results to test the actual evolution theory of stellar structure, but
we expect the ever-improving accuracy of the 
data will drive the theory to advance in new directions 
and eventually lead to a more thorough understanding of stellar rotation.  
Considering the fact that 
the internal angular velocity
 of the core of the red giants is theoretically expected
 to be higher compared to our
results, it remains still necessary to investigate more efficient
mechanisms of angular momentum transport acting on the appropriate timescales
during the different phases of the stellar evolution, before the red-giant phase.  We expect that the
measurements of rotational splittings for modes with low inertia will shed some light
on the above picture and on the question of the stellar angular momentum transport.
Although some preliminary tests \citep{beck2014} have shown that the use
of $l=2$ modes splittings cannot help to resolve internal rotation inside red giants, we
believe that a detailed analysis is required by
considering red giants at different evolutionary phases.

We conclude that it is reasonable to think that 
 this approach, proved to be very 
powerful in the case of the Sun, for which thousands of modes from low to high degree have been detected, can be well applied even to 
 small sets of only dipolar modes in red-giant~stars.

%%%%%%%%%%%%%%%%%%%%%%%%%%%%%%%%%%%%%%%%%%%%%%%%%%%%%%%%%%%%%%%%%%%%%%%%%%%
\begin{acknowledgements}
The authors thank very much the anonymous referee for his/her suggestion and comments, which gave  the opportunity to greatly improve the manuscript.

Funding for the Kepler mission is provided by NASA's Science Mission Directorate. We thank the entire Kepler team for the
development and operations of this outstanding mission.

WAD was supported by the Polish  NCN grant DEC-2012/05/B/ST9/03932.
SH acknowledges financial support from the Netherlands Scientific Organization (NWO).
DS acknowledges support from the Australian Research Council.
Funding for the Stellar Astrophysics Centre is provided by The Danish National Research Foundation (Grant agreement no.: DNRF106). The research is supported by the ASTERISK project (ASTERoseismic Investigations with SONG and Kepler) funded by the European Research Council (Grant agreement no.: 267864).
BM, PB and RAG acknowledge the ANR (Agence Nationale de la Recherche, France) program IDEE 
(n. ANR-12-BS05-0008) "Interaction Des \'Etoiles et des Exoplan\`etes".
AT is postdoctoral fellow of the Fund for Scientific Research (FWO), Flanders, Belgium.
The research leading to the presented results has received funding from the European Research Council under the European Community's Seventh Framework Programme (FP7/2007-2013) / ERC grant agreement no 338251 (StellarAges).
The work presented here is also based on
ground-based spectroscopic observations 
made with the Mercator Telescope, operated on the island of La Palma
by the Flemish Community, at the Spanish Observatorio del Roque de los
Muchachos of the Instituto de Astrofisica de Canarias.

MPDM is grateful to the Astrophysical Observatory of Catania for hosting her during the preparation of the present manuscript.

\end{acknowledgements}
%%%%%%%%%%%%%%%%%%%%%%%%%%%%%%%%%%%%%%%%%%%%%%%%%%%%%%%%%%%%%%%%%%%%%%%%%%%

\end{document}